\documentclass[12pt,letterpaper]{article}
\pdfoutput=1
\usepackage{amssymb}
\usepackage{amsmath}
\usepackage{amsfonts}
\usepackage{epsfig}
\usepackage{color}
\usepackage{hyperref}
\usepackage[normalem]{ulem}

\usepackage{bm}
\outer\def\beginsection#1\par{\medbreak\bigskip
      \message{#1}\leftline{\bf#1}\nobreak\medskip
\vskip-\parskip
      \noindent}

\newcommand{\eq}{\begin{equation}}
\newcommand{\eqx}{\end{equation}}
\newcommand{\eqn}{\begin{eqnarray}}
\newcommand{\eqnx}{\end{eqnarray}}
\newcommand{\bi}{\begin{itemize}}
\newcommand{\ei}{\end{itemize}}

\newcommand{\bA}{{\bf A}}

\def\Ord{{\cal O}}

\def\be{\begin{equation}}
\def\ee{\end{equation}}
\def\ba{\begin{eqnarray}}

\newcommand{\ea}[1]{\begin{align} #1 \end{align}}
\newcommand{\sea}[1]{\begin{subequations}\begin{align} #1 \end{align}\end{subequations}}

\newcommand{\seq}[1]{\begin{equation} \begin{split} #1 \end{split} \end{equation}}

\newcommand{\LN}[2]{\log | z_{#1}-z_{#2} |}
\newcommand{\zz}[2]{z_{#1}-z_{#2}}
\newcommand{\zbzb}[2]{\bar{z}_{#1}-\bar{z}_{#2}}

\def\bA{\bar{A}}
\def\bC{\bar{C}}

\usepackage[titles]{tocloft}

\usepackage{titlesec}
\titleformat*{\section}{\large  \bfseries }
\titleformat*{\subsection}{\normalsize  \bfseries }

\begin{document}

\begin{titlepage}
\hfill \hbox{NORDITA-2016-106}
\vskip 1.5cm
\vskip 1.0cm
\begin{center}
{\Large \bf
Soft behavior of a closed massless state in superstring and universality in the 
soft behavior of the dilaton}
 
\vskip 1.0cm {\large Paolo
Di Vecchia$^{a,b}$,
Raffaele Marotta$^{c}$, Matin Mojaza$^{d}$
} \\[0.7cm] 
{\it $^a$ The Niels Bohr Institute, University of Copenhagen, Blegdamsvej 17, \\
DK-2100 Copenhagen \O , Denmark}\\
{\it $^b$ Nordita, KTH Royal Institute of Technology and Stockholm 
University, \\Roslagstullsbacken 23, SE-10691 Stockholm, Sweden}\\[2mm]
{\it $^c$  Istituto Nazionale di Fisica Nucleare, Sezione di Napoli, Complesso \\
 Universitario di Monte S. Angelo ed. 6, via Cintia, 80126, Napoli, Italy}\\[2mm]
 {\it $^d$  Max-Planck-Institut f\"ur Gravitationsphysik, \\
Albert-Einstein-Institut, Am M\"uhlenberg 1, 14476 Potsdam, Germany}
\end{center}
\begin{abstract}
We consider the tree-level scattering amplitudes in the NS-NS (Neveu-Schwarz)
massless sector of closed superstrings in the case where one external state becomes soft. 
We compute the amplitudes generically for any number of dimensions and any 
number and kind of the massless closed states through the subsubleading order 
in the soft expansion. 
We show that, when the soft state is a graviton or a dilaton, the full result 
can be expressed as a soft theorem factorizing the amplitude in a soft and a 
hard part. This behavior is similar to what has previously been observed in field 
theory and in the bosonic string. Differently from the bosonic string, the 
supersymmetric soft theorem for the graviton has no string corrections at 
subsubleading order. The dilaton soft theorem, on the other hand, is found to 
be universally free of string corrections in any string theory.
\end{abstract}
\end{titlepage}

\tableofcontents
\section{Introduction and Results}
\label{intro}

In this work we consider the tree-level scattering amplitudes of 
massless states in the NS-NS (Neveu-Schwarz) sector of closed superstrings, 
and 
analyze the behavior of the amplitude when one of the external states, either a 
graviton, dilaton or a Kalb-Ramond field, becomes soft with respect to the other 
(hard) states.
This is a direct continuation of our earlier works in 
Ref.~\cite{DiVecchia:2015oba,DiVecchia:2016amo}, where we dealt with 
the same problem, but in the bosonic string only.
In Ref.~\cite{DiVecchia:2015oba} the soft behavior of a massless closed 
string was computed through subsubleading order in the soft momentum, 
when the hard states were all tachyons, and through subleading order, when
 the hard states were any other massless closed state in the bosonic string. 
These soft behaviors were shown for the graviton and dilaton to generically 
admit soft theorems with the factorizing soft part being equal to the known 
leading and subleading soft theorems of a graviton in pure field 
theory~\cite{Weinberg}, {and of a dilaton in string theory~\cite{SoftDilaton}}. 
The subsubleading soft behavior, when scattering on 
hard tachyons was found instead to have an additional factorizing piece, only 
relevant for a soft dilaton, as compared to the recently discovered subsubleading
 soft theorem for the graviton in 
field theory~\cite{Cachazo:2014fwa,Broedel:2014fsa, Bern:2014vva} 
(for complementary discussions, see also Ref.\cite{SoftGravity}).

As shown in \cite{Bern:2014vva}, the field theory soft theorem results for 
the graviton all follow by just imposing gauge invariance of the scattering amplitudes. This same analysis can be extended to also cover the dilaton collectively 
with the graviton, and as shown in Ref.~\cite{DiVecchia:2015jaq}, 
one indeed recovers the additional piece at 
subsubleading order, 
found explicitly in Ref.~\cite{DiVecchia:2015oba}, when scattering on hard 
tachyons, signalling universality of the soft theorem. In 
Ref.~\cite{DiVecchia:2016amo} we extended our analysis in the bosonic string
 by computing the subsubleading soft behavior of a massless closed state 
scattering on other hard massless closed states.  
The soft theorems for 
the graviton and dilaton were again uncovered, and it was shown that an 
additional factorizing soft operator proportional to the string slope $\alpha'$ 
appears at subsubleading order, when the soft state is a graviton, while the 
dilaton soft theorem remains equal to the field theory result 
of Ref.~\cite{DiVecchia:2015jaq}. 
By including the $\alpha'$ corrections in the three-point amplitude for 
massless states, this was shown to again follow from gauge invariance of 
the scattering amplitudes. 
While this shows that the graviton soft theorem at subsubleading order is not 
universal, but depends on higher-order operators in its effective action, it is 
intriguing to think that the soft behavior of the dilaton is universal in any 
theory through subsubleading order, signalling some underlying hidden 
symmetry. 
({For recent discussions on existing relations between broken symmetries of 
Lagrangians and soft theorems, see Refs.~\cite{Symmetries,DiVecchia:2015jaq}.}) 
Indeed, as shown in Ref.~\cite{DiVecchia:2015jaq}, the dilaton 
soft theorem does bare striking resemblance to the soft theorem of the 
Nambu-Goldstone boson of spontaneously broken conformal symmetry, 
which is universal through subsubleading order. 

In this work we are in fact 
going to confirm that the dilaton in superstrings obeys the same soft theorem
 as in the bosonic string through subsubleading order,
 when scattering on other massless closed states of
 the NS-NS sector of closed superstrings. The soft theorem of the graviton in 
superstrings, on the other hand, does not have any string corrections, in 
contrast to the bosonic string. {Furthermore, we will show by using gauge invariance 
that also in the heterotic string, the dilaton soft operator has no 
string corrections at subsubleading order. 
In conclusion, we find that the dilaton soft operator is tree-level universal 
through subsubleading order in all string theories, and in particular, does not 
contain string corrections, making it the same as in field theory. 
We leave out in this work any discussions on 
the soft behavior of the Kalb-Ramond field. We plan to discuss it in a future 
work, collecting also our previous results in the bosonic string.

{The subject of this work has seen tremendous progress in recent years, both
 in field theory and in string theories. We restrict ourselves here 
to referring only to the more related string theory papers~\cite{softstring}, 
while the interested reader is invited to look up our recent 
paper~\cite{DiVecchia:2016amo} for a brief count of the progress also 
on the field theory side, including the relevant references.

Let us summarize the results of this work, while at the same time introducing our notation: 
The $n$-point tree-level scattering amplitudes, $M_n$, of closed massless 
superstrings can generically be written as a convolution of a bosonic part, 
$M_n^b$, 
with a supersymmetric part, $M_n^s$, as follows:
\ea{
M_n = M_n^b \ast M_n^s \, ,
}
The expressions for bosonic and supersymmetric parts of the $n$-point 
supersymmetric 
string amplitude are defined by: 
\seq{
M_n^b =  &\, \frac{8\pi}{\alpha'}\left (\frac{\kappa_D}{2\pi}\right )^{n-2}  
\int \frac{ \prod_{i=1}^{n} d^2 z_i}{dV_{abc} |z_1  - z_2|^2}  
{\prod_{i=1}^{2}  d \theta_i \theta_i  
\prod_{i=1}^{2} d \bar{\theta}_i \bar{\theta}_i}
\prod_{i=3}^{n} d \theta_i  
\prod_{i=1}^{n} d \varphi_i   
\prod_{i=3}^{n}   d {\bar{\theta}}_i  
\prod_{i=1}^n d {\bar{\varphi}}_i \\
&  \times  \prod_{i < j} | z_i - z_j |^{\alpha' k_i \cdot k_j}\, \exp \left[ \frac{1}{2} 
\sum_{i\neq j} \frac{C_i \cdot C_j}{(\zz{i}{j})^2} + \sqrt{\frac{\alpha'}{2}}
\sum_{i\neq j} \frac{C_i \cdot k_j}{\zz{i}{j}} + \text{c.c.} \right ] \, ,
}
\ea{
{M}_n^s = \exp \left [ - \frac{1}{2} \sum_{i\neq j} \frac{A_i \cdot A_j}{\zz{i}{j}} + 
\text{c.c.} \right] \, .\label{1.5}
}
where $\kappa_D$ is the $D$-dimensional Newton's constant,  $dV_{abc}$ 
is the volume of the M\"obius group, $z_i$ are the Koba-Nielsen variables,
$\varphi_i$ and $\theta_i$ are Grassmannian integration variables,
and  we have introduced the following \emph{superkinematical} quantities:
\ea{
 A_i^\mu = \varphi_i \epsilon_i^\mu  +\sqrt{\frac{\alpha'}{2}} \theta_i k_i^\mu
~~;
 ~~ C_i^\mu = \varphi_i \theta_i  \epsilon_i^\mu \, ,\label{1.4}
}
where $\epsilon_i^\mu$ and $k_i^\mu$ are respectively the holomorphic 
polarization vector and momentum of the state $i$, and $\alpha'$ is the string 
slope.}

Apart from the integration measure, $M_n^b$ is equivalent at the integrand 
level to the same amplitude in the bosonic string; the integrands, in fact, 
become equal if one makes the identification 
 $\theta_i \epsilon_i \to \epsilon_i$ and remembers
that, after this substitution,  $\epsilon_i$ has become a Grassmann variable.
The difference in the measure between $M_n^b$ and the bosonic string 
amplitude
is only the presence in $M_n^b$ of the integrals over the Grassmann 
variables $\theta_i$, $\bar{\theta}_i$,  and the additional factor $\prod_{i=1}^2 
\theta_i \bar{\theta}_i/|z_1-z_2|^2$ coming from the correlator of 
the superghosts.
The latter factor in the measure effectively kills any term 
involving $\theta_1, \theta_2, \bar{\theta}_1, \bar{\theta}_2$, 
which readily follows from an expansion of the exponentials and 
an integration over those variables. We will not need to use this property, and 
thus leave the integrand as it is.

When considering an amplitude with an additional  state, which is soft, it is 
useful to factorize the string amplitude at the integrand level into a 
soft part $S$ and a hard part as follows:
\ea{
M_{n+1} = M_n \ast S
}
where $M_n$ is the full superstring amplitude of $n$ closed massless states, 
and $S$ is a function that when convoluted with the integral expression 
for $M_n$ provides the additional soft state involved in the amplitude. 
The function $S$ can be further decomposed into its bosonic part and 
supersymmetric part as follows:
\ea{
S = S_b+S_s+\bar{S}_s \, ,
}
where $S_b$ is the bosonic part and $S_s+\bar{S}_s$ is the 
supersymmetric part, with $\bar{S}_s$ being the complex conjugate of $S_s$. 
{This decomposition is useful, since 
 the bosonic part can be related to the soft function in the bosonic string. 
In fact, after the identification $\theta_i\epsilon_i\to  \epsilon_i$, the 
bosonic soft function $S_b$ is the same as the one 
given for the bosonic string in Ref.~\cite{DiVecchia:2015oba,DiVecchia:2016amo},%
\footnote{The Grassmanian variables $\varphi_i$ are equivalent to those of Ref.~\cite{DiVecchia:2015oba,DiVecchia:2016amo} denoted therein by $\theta_i$.}
which was computed therein through $\Ord(q)$,
where $q$ is the momentum of the soft state.
In this work we need therefore only to consider the additional contributions from the supersymmetric states, described by $S_s+\bar{S}_s$. We have computed this function through $\Ord(q)$, and our result reads:}
 \ea{
&S_s+\bar{S}_s =
\kappa_D \epsilon_\mu \bar{\epsilon}_\nu  
 \sum_{i\neq j}  \Bigg \{
\frac{q_\rho}{(k_i\cdot q)} \frac{ \bar{A}_i^{[\rho}\bar{A}_j^{\nu]} 
k_i^\mu}{(\bar{z}_i-\bar{z}_j)}
 +
 q_\rho\left(\frac{\alpha'}{2}\right)^{\frac{3}{2}}
\frac{q\cdot k_j \bC_i^{[\rho,}k_i^{\nu]}}{\zbzb{i}{j}} 
\left (\frac{k_i^\mu}{q\cdot k_i} - \frac{k_j^\mu}{q\cdot k_j}\right ) 
 \nonumber \\
 &
 +q_\rho\sqrt{\frac{\alpha'}{2}}
\frac{\bA_{\{i ,}^\rho \bA_{j\} }^\nu}{\zbzb{i}{j}} \sum_{l\neq i} 
\Bigg [
\frac{q\cdot k_l}{q\cdot k_i} \left (\frac{C_i^\mu}{\zz{i}{l}} +
\sqrt{\frac{\alpha'}{2}} k_i^\mu \LN{i}{l}^2 \right )
 + 
\left (\frac{C_l^\mu}{\zz{i}{l}} - \sqrt{\frac{\alpha'}{2}}k_l^\mu \LN{i}{l}^2 \right )  
\Bigg ]
\nonumber \\
&
+q_\rho q_\sigma \Bigg[
\left (
\frac{1}{2}
A_{\{i,}^\sigma A_{j\}}^\mu -\sqrt{\frac{\alpha'}{2}} 
C_{\{i,}^\sigma k_{j\}}^\mu\right ) 
\sum_{l\neq i}\frac{\bA_{\{i ,}^\rho \bA_{l\} }^\nu}{q\cdot 
k_i (\zz{i}{j})(\zbzb{i}{l})} 
-\frac{\alpha'}{2}
\frac{\bC_i^{[\sigma,}k_i^{\nu]} \bC_j^\rho}{(\zbzb{i}{j})^2}
\left (\frac{k_j^\mu}{q\cdot k_j} - \frac{k_i^\mu}{q\cdot k_i}\right ) 
\nonumber \\
&
 -\sqrt{\frac{\alpha'}{2}} \sum_{l\neq i,j} \frac{k_i^\mu 
\left ( \bC_j^\sigma \bA_{\{ i,}^\rho \bA_{l\}}^\nu +\frac{1}{2} \bC_i^\sigma 
\bA_{\{ j,}^\rho \bA_{l\}}^\nu \right )}{q\cdot k_i (\zbzb{i}{j})(\zbzb{i}{l})}
- \sum_{l\neq i} \frac{C_{[i,}^\sigma C_{j]}^\mu \bA_{\{i ,}^\rho 
\bA_{l\} }^\nu}{q\cdot k_i(\zz{i}{j})^2 (\zbzb{i}{l})}  
\Bigg] \Bigg\} + \text{c.c.} + \Ord(q^2) \, ,
}
where the brackets and curly-brackets in the indices denote 
commutation and anticommutation of the indices, e.g.:
\seq{
 C_i^{[\rho ,} k_i^{\nu]} &\equiv C_i^\rho k_i^\nu - C_i^\nu k_i^\rho \\
 A_{\{i}^\mu A_{j\}}^\nu &\equiv A_i^\mu A_j^\nu + A_j^\mu A_i^\nu
 }
The above expression starts at $\Ord(q^0)$. This means that only the 
bosonic part contributes to the amplitude at $\Ord(q^{-1})$, and is thus 
responsible for the Weinberg graviton soft theorem. When the above 
expression is projected onto a soft state which is symmetric in its 
polarization, i.e. a graviton or a dilaton, we will show that the explicit 
results above can be reproduced by the following soft theorem:
 \ea{
 (M_{n+1})_S = (\hat{S}^{(-1)} + \hat{S}^{(0)}+\hat{S}^{(1)}) M_n + \Ord(q^2)
 }
 where the subscript $S$ denotes that the soft state must be 
symmetrically polarized, and
 \sea{
 \hat{S}^{(-1)}  &= \kappa_D \, \epsilon_{\mu \nu}^S \sum_{i=1} 
\frac{k_i^\mu k_i^\nu}{k_i \cdot q} \, ,
 \\
 \hat{S}^{(0)}
& =
 - i \kappa_D \epsilon_{\mu \nu}^S \sum_{i=1}^n \frac{q_\rho 
k_i^\nu  J_i^{\mu \rho}}{k_i \cdot q} \, ,
 \\
 \hat{S}^{(1)}
& =
- \kappa_D \frac{\epsilon_{\mu \nu}^S}{2} \sum_{i=1}^n
\left (
\frac{q_\rho J_i^{\mu \rho} q_\sigma J_i^{\nu \sigma}}{k_i \cdot q}
+ \frac{q^\mu \eta^{\nu \rho} q^\sigma 
+q^\mu \eta^{\nu \sigma} q^\rho 
- \eta^{\mu\nu} q^\sigma q^\rho }{k_i \cdot q}
\mathbf{A}_{i\rho \sigma} 
\right ) \, , 
\label{sumhatS1}
 }
 and where $\epsilon_{\mu\nu}^{S}=
\frac{1}{2}(\epsilon_\mu\bar{\epsilon}_\nu+
\epsilon_\nu\bar{\epsilon}_\mu)$, $J_i$ is the total angular momentum operator,
\ea{
J_i^{\mu \nu} &= L_i^{\mu \nu} + S_i^{\mu \nu} + {\bar{S}}^{\mu \nu}_i \ ,
\label{Ji}
 \\[2mm]
 L_i^{\mu\nu} =i\left( k_i^\mu\frac{\partial }{\partial k_{i\nu}} -
k_i^\nu\frac{\partial }{\partial k_{i\mu}}\right) \, , \ 
S_i^{\mu\nu}&=i\left( \epsilon_i^\mu\frac{\partial }{\partial 
\epsilon_{i\nu}} -\epsilon_i^\nu\frac{\partial }{\partial 
\epsilon_{i\mu}}\right) \, , \  
{\bar{S}}^{\mu\nu}_i=i\left( {\bar{\epsilon}}_i^\mu
\frac{\partial }{\partial {\bar{\epsilon}}_{i\nu}} -
{\bar{\epsilon}}_i^\nu\frac{\partial }{\partial {\bar{\epsilon}}_{i\mu}}\right)  \, ,
\nonumber
}
and $\mathbf{A}_i$ is an operator:
\ea{
\mathbf{A}_{i\rho \sigma}= k_{i\rho}\frac{\partial}{\partial k_i^\sigma}+ 
\Pi_{i\rho \sigma}
\, , \quad
\Pi_{i\rho \sigma}= \epsilon_{i\rho} \frac{\partial}{\partial 
\epsilon_i^\sigma }+\bar{\epsilon}_{i\rho} 
\frac{\partial}{\partial \bar{\epsilon}_i^\sigma }
\, 
\label{Ai}
}
that acts covariantly on  the superkinematical variables, i.e.
\ea{
\mathbf{A}_i^{\mu \rho} A_j^\sigma =  
\delta_{ij} \eta^{\sigma \rho} A_i^\mu \, , \quad
\mathbf{A}_i^{\mu \rho} C_j^\sigma =  
\delta_{ij} \eta^{\sigma \rho} C_i^\mu \, .
\label{Aaction}
}
It follows that $J_i^{\mu \nu}$ is also covariant acting on these variables, since
\ea{
J_i^{\mu \nu} = i (\mathbf{A}_i^{\mu \nu} - \mathbf{A}_i^{\nu \mu} ) \, .
}
{In contrast, we notice that the 
subsubleading soft operators obtained in the bosonic string~\cite{DiVecchia:2016amo}: 
\ea{
 &\hat{S}_{bos}^{(1)} = \hat{S}^{(1)} +
\frac{\alpha'}{2} \kappa_D\, \epsilon_{\mu \nu}^S\sum_{i=1}^n 
\left (q^\sigma k_{i}^\nu \eta^{\rho \mu}+q^\rho k_{i}^\mu 
\eta^{\sigma \nu} - \eta^{\rho\mu}\eta^{\sigma \nu} (k_i \cdot q) - q^\rho 
q^\sigma \frac{k_{i}^\mu k_{i}^\nu}{q\cdot k_i} \right )
\Pi_{i\rho \sigma} \, ,
}
and, as we will show in this work, also in heterotic string:
\ea{
\hat{S}_{het}^{(1)} = \hat{S}^{(1)} +
\frac{\alpha'}{2} \kappa_D\, \epsilon_{\mu \nu}^S\sum_{i=1}^n 
\left (q^\sigma k_{i}^\nu \eta^{\rho \mu}+q^\rho k_{i}^\mu 
\eta^{\sigma \nu} - \eta^{\rho\mu}\eta^{\sigma \nu} (k_i \cdot q) - q^\rho 
q^\sigma \frac{k_{i}^\mu k_{i}^\nu}{q\cdot k_i} \right )
\epsilon_{i\rho} \frac{\partial}{\partial \epsilon_i^\sigma}
\, ,
}
differ from Eq.~\eqref{sumhatS1} by terms due to string corrections.}
These additional parts, proportional to $\alpha'$, do not act 
covariantly on the superkinematical variables, and thus are not
supersymmetric operators. We have consistently found that they only 
appear in the bosonic and in the heterotic 
string. They furthermore vanish when projected 
onto the dilaton state. 
Therefore the subsubleading soft operator in Eq.~\eqref{sumhatS1} is, nevertheless, universally valid for the dilaton in the bosonic string, in superstrings,
in the heterotic string, and in field theory~\cite{DiVecchia:2015jaq}.

The paper is organized as follows:
{In Sec.~\ref{TheAmplitude} we review the superstring amplitude of $n+1$ 
massless closed states and rewrite it in a convenient form for computing 
its  behavior in the limit  where one of the external states becomes soft with respect to the momenta of the other $n$ external states.}
Here we also introduce our notation. Then in Sec.~\ref{SoftExpansion} 
we show the calculational details of the soft part of the amplitude and 
provide our explicit results.
In Sec.~\ref{SoftAction} we demonstrate that the explicit results for the 
graviton and the dilaton can be expressed equally as a soft theorem, 
where the soft part is provided by the action of an operator acting 
on the lower point amplitude involving only the $n$ external hard states. 
We furthermore explicitly show how the supersymmetric part of the amplitude 
cancels the purely bosonic string corrections to the amplitude at the 
subsubleading order, found in Ref.~\cite{DiVecchia:2016amo}.  
In Sec.~\ref{heterotic}, using
gauge invariance, we compute the string corrections in the heterotic string
and we show that they do not contribute to the dilaton soft behavior.
Finally, Sec.~\ref{Conclusions} offers our conclusions and remarks. 
An appendix is additionally provided for the details of the calculation 
in Sec.~\ref{SoftAction}.

\section{Amplitude of one soft and $n$ massless closed superstrings}
\label{TheAmplitude}
\setcounter{equation}{0}

In this section, we review the closed superstring amplitude and rewrite it in a 
convenient form, when one particle is soft, which allows us to directly express 
the results using the calculations already done in the bosonic string 
in Ref.~\cite{DiVecchia:2015oba,DiVecchia:2016amo}.

The massless closed superstring vertex, in the $(-1,-1)$ and $(0,\,0)$ 
pictures,  is given by the compact expression:
\begin{eqnarray}
V^{(-p,\,-p)}=\frac{\kappa_D}{2\pi}\int d\theta~\theta^p~ V^{(p)}(z,\,\theta;k) 
\int d\bar{\theta} ~\bar{\theta}^{p} ~  \bar{V}^{(p)}(\bar{z},\,\bar{\theta}; k) \, ,
\end{eqnarray} 
with
\ea{
V^{(p)}(z,\,\theta;k)=  e^{-p\phi(z)}~\epsilon_\mu~ DX^\mu\, 
e^{i\sqrt{\frac{\alpha'}{2}} k\cdot X(z,\,\theta)}
\, ,
}
where $\theta$ and $\bar{\theta}$ are Grassmannian variables,
$\epsilon_\mu\bar{\epsilon}_\nu = \epsilon_{\mu\nu}$ is the polarization
 of the massless state,
and the superfield notation is given by \ea{
 X^{\mu}  (z, \theta) &\equiv x^{\mu}(z) + \theta \psi^{\mu}(z) \, , \quad
D \equiv \frac{\partial}{\partial \theta} + \theta \frac{\partial}{\partial z} \, . 
\label{3}
}
The relevant expectation values for massless amplitudes are:
\seq{
\langle X^{\mu} (z_1, \theta_1) X^{\nu} (z_2, \theta_2) \rangle &= - 
\eta^{\mu \nu} \log(z_1 -z_2 - \theta_1 \theta_2) \, , \\
\langle e^{-\phi(z_1)}\,e^{-\phi(z_2)}\rangle &=\frac{1}{z_1-z_2} \, .
\label{5}
}
The amplitude of $n+1$ massless states in closed superstring can be written 
as:
\ea{
M_{n+1} =&
\frac{8\pi}{\alpha'}\left (\frac{\kappa_D}{2\pi}\right )^{n-1}
 \int \frac{d^2 z \prod_{i=1}^{n} d^2 z_i d\theta d\bar{\theta} }{dV_{abc} 
|z_1  - z_2|^2}\bigg[\prod_{i=1}^2d\theta_i\theta_i \prod_{i=3}^{n} 
d \theta_i\bigg]\bigg[\prod_{i=1}^2d\bar{\theta}_i\bar{\theta}_i 
\prod_{i=3}^{n} d {\bar{\theta}}_i\bigg]  \nonumber \\
&  \times  \langle 0 | 
\int d \varphi \,\,e^{ i \left( \varphi \epsilon  D X (z, \theta) + 
\sqrt{\frac{\alpha'}{2}}q  X (z, \theta)  \right)}
\prod_{i=1}^{n} \left( \int d \varphi_i  \,\,e^{ i \left( \varphi_i 
\epsilon_i  D_i X (z_i, \theta_i) + K_i  X (z_i, \theta_i)  \right)} \right) 
| 0\rangle \nonumber \\
&  \times  \langle 0 | 
\int d {\bar{\varphi}} \,\,e^{ i \left( {\bar{\varphi}} {\bar{\epsilon}} 
 {\bar{D}} X ({\bar{z}}, {\bar{\theta}}) + \sqrt{\frac{\alpha'}{2}}q  
X ({\bar{z}}, {\bar{\theta}})  \right)}
\prod_{i=1}^{n} \left( \int d {\bar{\varphi}}_i  \,\,e^{ i 
\left( {\bar{\varphi}}_i {\bar{\epsilon}}_i  {\bar{D}}_i X ({\bar{z}}_i, {\bar{\theta}}_i) + K_i  X ({\bar{z}}_i, {\bar{\theta}}_i)  \right)} \right) | 0\rangle \, ,
\label{super1}
}
where new Grassmanian variables $(\varphi, \varphi_i, \bar{\varphi}, 
\bar{\varphi}_i)$  are introduced,
and $dV_{abc}$ is the volume of the M\"obius group. 
{The states with the indices $1$ and $2$ are in the $(-1, -1)$
picture, while the others are in the $(0,0)$ picture. This effectively means that,
in the expressions for the integrands that follow, terms involving $\theta_1, \theta_2, {\bar{\theta}}_1, {\bar{\theta}}_2$ can be equated to zero because of the overall integration measures $\int d \theta_i \theta_i$ and  $\int d \bar{\theta}_i \bar{\theta}_i$ for $i=1,2$.
Since this choice could have been made for any two of the $n$ states, we will not explicitly impose these zero conditions in the expressions that follow.
 }

The $n+1$ point amplitude,  with the help of the correlation functions
 written in Eq.~(\ref{5}) and after having integrated over the variables 
$\theta$ and $\bar{\theta}$, {reduces to an expression which can be
factorized at the integrand level as follows:}
\begin{eqnarray}
M_{n+1}=M_n*S \, ,
\end{eqnarray}
{where by $*$ a convolution integral is understood, and  the two parts 
$M_n$ and $S$ 
can be conveniently expressed in terms of the 
 superkinematical quantities:
\begin{eqnarray}
 A_i^\mu = \varphi_i \epsilon_i^\mu  +\sqrt{\frac{\alpha'}{2}} \theta_i 
k_i^\mu~~;
 ~~ C_i^\mu = \varphi_i \theta_i  \epsilon_i^\mu \, ,
 \label{AiBiCi}
 \end{eqnarray}
  such that
\seq{
 M_{n} = &\, \frac{8\pi}{\alpha'}\left (\frac{\kappa_D}{2\pi}\right )^{n-2}  
\int \frac{ \prod_{i=1}^{n} d^2 z_i}{dV_{abc} |z_1  - z_2|^2} 
{\prod_{i=1}^{2}  d \theta_i \theta_i  
\prod_{i=1}^{2} d \bar{\theta}_i \bar{\theta}_i}
\prod_{i=3}^{n} d \theta_i  
\prod_{i=1}^{n} d \varphi_i   
\prod_{i=3}^{n}   d {\bar{\theta}}_i  
\prod_{i=1}^n d {\bar{\varphi}}_i 
\\
&   \prod_{i < j} | z_i - z_j |^{\alpha' k_i k_j}\, \exp \left[ \frac{1}{2} 
\sum_{i\neq j} \frac{C_i \cdot C_j}{(\zz{i}{j})^2} +\sqrt{\frac{\alpha'}{2}} 
\sum_{i\neq j} \frac{C_i \cdot k_j}{\zz{i}{j}} - \frac{1}{2} \sum_{i\neq j} 
\frac{A_i \cdot A_j}{\zz{i}{j}} \right] 
 \\
& \hspace{22.3mm} \quad \times \exp \left[ \frac{1}{2} \sum_{i\neq j} 
\frac{\bC_i \cdot \bC_j}{(\zbzb{i}{j})^2} + \sqrt{\frac{\alpha'}{2}}
\sum_{i\neq j} \frac{\bC_i \cdot k_j}{\zbzb{i}{j}} - \frac{1}{2} \sum_{i\neq j} 
\frac{\bA_i \cdot \bA_j}{\zbzb{i}{j}} \right]\, ,
\label{Mncompact}
}
}
while for convenience we express  $S$ as a sum of three terms
\begin{eqnarray}
S\equiv S_b+S_s+\bar{S}_s\label{defS}\, ,
\end{eqnarray}
where $S_b$ is the purely bosonic part, 
{which is simply equal to the similar expression in the bosonic string after
 identifying $\theta_i \epsilon_i \to  \epsilon_i$ (whereby 
$\epsilon_i$ becomes a Grassmann variable)} and is given by:
\footnote{For comparison with the expressions in Ref.~\cite{DiVecchia:2015oba,DiVecchia:2016amo} we notice that the variables here denoted by $\varphi_i$ are equivalent to the variables denoted by $\theta_i$ in those papers.} 
\seq{
 S_b= &\,\frac{\kappa_D}{2\pi} \int d^2 z \prod_{l=1}^{n}
 | z - z_l |^{ \alpha' q k_l} 
\,\prod_{l=1}^{n} {\rm exp}\left [ -\sqrt{\frac{\alpha'}{2}}\frac{q \cdot C_l}{z-z_l}-  
\sqrt{\frac{\alpha'}{2}}\frac{q \cdot \bar{C}_l}{{\bar{z}}-{\bar{z}}_l} \right] 
 \\
& \times 
\left( \sum_{i=1}^{n} \frac{\epsilon \cdot C_i}{(z-z_i)^2} +  
\sum_{i=1}^{n}\sqrt{\frac{\alpha'}{2}} \frac{\epsilon \cdot k_i}{z-z_i} \right)
\left( \sum_{j=1}^{n} \frac{\bar{\epsilon} \cdot 
{\bar{C}}_j}{({\bar{z}}-{\bar{z}}_j)^2} + \sum_{j=1}^{n}\sqrt{\frac{\alpha'}{2}} 
\frac{{\bar{\epsilon}} \cdot k_j}{{\bar{z}}-{\bar{z}}_j} \right)
\, ,
}
and $S_s$ and $\bar{S}_s$ are the complex conjugates of each other and 
they provide the contributions from the additional supersymmetric states. 
They are given by
\seq{
\bar{S}_s = &\, \frac{\kappa_D}{2\pi} \int d^2 z\prod_{l=1}^{n}
 | z - z_l |^{ \alpha' q k_l} 
\,\prod_{l=1}^{n}  {\rm exp}\left [ -\sqrt{\frac{\alpha'}{2}}\frac{q \cdot C_l}{z-z_l}-  
\sqrt{\frac{\alpha'}{2}}\frac{q \cdot \bar{C}_l}{{\bar{z}}-{\bar{z}}_l} \right]  \\
& \times \left[ \frac{1}{2}\sum_{i=1}^{n} \sqrt{\frac{\alpha'}{2}}
\frac{q \cdot A_i}{z-z_i} \sum_{j=1}^{n} \frac{\epsilon \cdot A_j}{z-z_j}
 \sum_{l=1}^{n}\sqrt{\frac{\alpha'}{2}} \frac{q \cdot
 {\bar{A}}_l}{{\bar{z}}-{\bar{z}}_l} \sum_{m=1}^{n} \frac{\bar{\epsilon} 
\cdot{\bar{A}}_m}{{\bar{z}}-{\bar{z}}_m} \right.
 \\
 &\left.
 + \left (\sum_{i=1}^{n} \frac{\epsilon \cdot C_i}{(z-z_i)^2} +  
\sum_{i=1}^{n}\sqrt{\frac{\alpha'}{2}} \frac{\epsilon \cdot k_i}{z-z_i} 
\right ) \sum_{j=1}^{n}\sqrt{\frac{\alpha'}{2}} \frac{q 
\cdot {\bar{A}}_j}{{\bar{z}}-{\bar{z}}_j} \sum_{l=1}^{n} \frac{\bar{\epsilon} 
\cdot{\bar{A}}_l}{{\bar{z}}-{\bar{z}}_l} 
  \right] \, ,
}
and  ${S}_s$ is given by  the complex conjugate of this expression,
where complex conjugation sends $z_i \to \bar{z}_i$, $\epsilon_i^\mu \to \bar{\epsilon}_i^\mu$, $\theta_i \to \bar{\theta}_i$, and $\varphi_i \to \bar{\varphi}_i$, while the momenta $k_i$ are left invariant. 
{The superkinematical quantities $A_i^\mu$ and $C_i^\mu$
 are respectively anticommuting and commuting kinematic factors.} 
Furthermore, 
 since $\varphi_i^2 = \theta_i^2 = 0$, they obey the following useful identities:
 \ea{
 A_i^\mu A_i^\nu = \sqrt{\frac{\alpha'}{2}}C_i^{[\mu ,} k_i^{\nu]} \, ,  \quad
  C_i^\mu C_i^\nu = A_i^\mu C_i^\nu = 0 \, ,
 \label{nulrelation}
 }
 where we have used the notation $C_i^{[\mu ,} k_i^{\nu]} \equiv C_i^\mu 
k_i^\nu - C_i^\nu k_i^\mu$. This antisymmetrizing notation will be used throughout 
this paper. Furthermore an equivalent notation will be used with curly brackets for 
denoting symmetrization.

Let us remark that  $M_n$ can be decomposed in a bosonic and a 
supersymmetric part as well, as follows:
\ea{
M_n = M_n^b \ast {M}_n^s \, ,
\label{1.16}
}
where the first part yields the complete bosonic case and is given by 
\seq{
M_n^b =  &\, \frac{8\pi}{\alpha'}\left (\frac{\kappa_D}{2\pi}\right )^{n-2}  
\int \frac{ \prod_{i=1}^{n} d^2 z_i}{dV_{abc} |z_1  - z_2|^2} {\prod_{i=1}^2 d\theta_i\theta_i} \prod_{i=3}^{n} 
 d \theta_i  \prod_{i=1}^{n} d \varphi_i  {\prod_{i=1}^2 d\bar{\theta}_i\bar{\theta}_i} \prod_{i=3}^{n}  
d {\bar{\theta}}_i  \prod_{i=1}^n  d {\bar{\varphi}}_i \\
&   \prod_{i < j} | z_i - z_j |^{\alpha' k_i k_j}\, \exp \left[ \frac{1}{2}
 \sum_{i\neq j} \frac{C_i \cdot C_j}{(\zz{i}{j})^2} + \sqrt{\frac{\alpha'}{2}}
\sum_{i\neq j} \frac{C_i \cdot k_j}{\zz{i}{j}} + \text{c.c.} \right ] \, ,
\label{Mnb}
} 
and the second part gives the supplement of the additional superstring 
states and reads
\ea{
{M}_n^s = \exp \left [ - \frac{1}{2} \sum_{i\neq j} \frac{A_i \cdot 
A_j}{\zz{i}{j}} + \text{c.c.} \right] \, .
\label{Mns}
}
    
 \section{Soft expansion through subsubleading order}
 \label{SoftExpansion}
\setcounter{equation}{0}
 
The integral $S_b$  has been computed 
through subsubleading order in $q$, 
that is through $\Ord(q)$, in 
Refs.~\cite{DiVecchia:2015oba,DiVecchia:2016amo}. Thus for this work we 
only need to consider the other parts of $S$, i.e. ${S}_s$ and $\bar{S}_s$, 
where the latter can be conveniently written in the following compact form:
\ea{
\bar{S}_s = \,\sqrt{\frac{\alpha'}{2}} \kappa_D \epsilon_\mu \bar{\epsilon}_\nu 
&\Bigg \{
q_\rho\sum_{i,j,l=1} \bA_j^\rho \bA_l^\nu \left ( C_i^\mu I_{ii}^{jl} +
\sqrt{\frac{\alpha'}{2}} k_i^\mu I_i^{jl} \right ) \nonumber \\
&
+\sqrt{\frac{\alpha'}{2}} \, q_\rho q_\sigma \sum_{i,j,l,m=1} 
\bA_l^\sigma \bA_m^\nu \Bigg [
\left (\frac{1}{2}A_i^\rho A_j^\mu -\sqrt{\frac{\alpha'}{2}} 
C_i^\rho k_j^\mu \right )  I_{ij}^{lm}
\nonumber \\
&
-
C_i^\rho C_j^\mu I_{ijj}^{lm} 
- C_i^\mu \bC_j^\rho I_{ii}^{jlm} - \sqrt{\frac{\alpha'}{2}}k_i^\mu 
\bC_j^\rho I_{i}^{jlm} \Bigg]
\Bigg \} \, ,
\label{Ss}
}
where all the integrals involved in the calculus  of the amplitude are represented 
as:
\ea{
I_{i_1 i_2 \ldots}^{j_1 j_2 \ldots} =
\int \frac{d^2 z}{2 \pi} \frac{\prod_{l = 1}^n |z-z_l|^{\alpha' qk_l}}{
(z-z_{i_1})(z-z_{i_2}) \cdots (\bar{z}-\bar{z}_{j_1}) (\bar{z}-\bar{z}_{j_2}) 
\cdots } \ .
\label{GeneralIntegralintext}
}
Notice that according to Eq.~\eqref{nulrelation} the term involving 
$C_i^\rho C_j^\mu$ vanishes for $i=j$, and that the terms
 involving $\bC_j^\rho$ vanish for $j=l,m$.
It turns out that all integrals involved in the calculation have already 
been computed in Ref.~\cite{DiVecchia:2016amo}, and they are all 
obtained from two master integrals,  $I_i^i$ and  $I_i^j$, through an 
iteratively use of the identities:
\begin{eqnarray}
 I_{ii}^{j}=\frac{1}{1-\frac{\alpha'}{2} (q k_i)}\partial_{z_i} I_{i}^j\label{1.22}
 \end{eqnarray}
 valid even for $i=j$ and 
 \begin{eqnarray}
 I^{j_1j_2\dots}_{i_1i_2\dots} =
\frac{I_{i_1\dots} ^{j_1j_2\dots}- I_{i_2\dots} ^{j_1j_2\dots}}{z_{i_1}-z_{i_2}}=
 \frac{I_{i_1 \dots}^{j_1 \dots}-I_{i_1\dots}^{j_2 \dots}-
I_{i_2\dots}^{j_1\dots}+I_{i_2\dots}^{j_2\dots}}{(z_{i_1}-z_{i_2})(
\bar{z}_{j_1}-\bar{z}_{j_2})}=\dots\label{1.23}
 \end{eqnarray}
The explicit expressions of the master integrals 
are~\cite{DiVecchia:2015oba,DiVecchia:2016amo}:
\ea{
I_i^i = & 
 \frac{2}{\alpha'(k_i q)}  
\left( 1 +\alpha'  \sum_{j \neq i} (k_j q) \log |z_i - z_j|  
 + \frac{(\alpha')^2}{2} 
\sum_{j \neq i} \sum_{k \neq i} (k_j q) (k_k q)  \log|z_i -z_j| \log |z_i - z_k| 
\right) 
 \nonumber \\
&+(\alpha')^2 \sum_{j \neq i} (k_j q) \log^2 |z_i - z_j| + \log \Lambda^2 + 
\Ord(q^2)
\, ,
\label{A}
 \\[5mm]
I_i^j = &
\sum_{m\neq i,j}\frac{\alpha'(qk_{m})}{2}\left ({\rm Li}_2\left( \frac{\bar{z}_i-
\bar{z}_m}{\bar{z}_i-\bar{z}_j}\right)-{\rm Li}_2\left(\frac{z_i-z_m}{z_i-z_j 
}\right)
-2\log\frac{\bar{z}_m-\bar{z}_j}{\bar{z}_i-\bar{z}_j}\log
\frac{ |z_i-z_j|}{|z_i-z_m|}
\right )
\nonumber\\
&- \log|z_i-z_j|^2+\log\Lambda^2  + \Ord(q^2) \, ,
\label{IIj}
}
with $\Lambda$ a cut off  that cancels  in the final expression of the 
amplitude. 
The notation of two momenta in a round bracket is hereafter used to denote
 $(k_j q) \equiv k_j \cdot q$.
It is worthwhile to notice that only $I_i^i$ shows a pole in the soft 
momentum and therefore the integrals $I_{i_1i_2\dots}^{j_1j_2\dots}$ 
can yield a term of $\Ord(q^{-1})$ only if one of its lower indices is equal 
to one of the upper ones.

To derive $\bar{S}_s$ through subsubleading order, let us first notice that 
with the integrand explicitly containing a factor of $q$, the leading part can 
only be of $\Ord(q^0)$, and therefore the entire $\Ord(q^{-1})$ terms are 
produced by the bosonic part only. Next, to obtain the terms of order $q^0$ 
and $q$, we notice by inspection of Eq.~(\ref{Ss}) that the integrals $I_{ii}^{jl}$
 and $ I_i^{jl}$ must be equated through the $\Ord(q^0)$,  while for all 
other integrals only the leading $q^{-1}$ order is relevant.
The integral $I_{ii}^{jlm}$, only relevant at $\Ord(q^{-1})$, does not contribute, 
since by having two lower indices equal it cannot be divergent in the soft 
momentum. For the same reason, all the other integrals which are only 
relevant at $\Ord(q^{-1})$ contribute only when one of the indices $l$ or $m$ 
is equal to $i$ or $j$.

The complete expression through $\Ord(q)$ of $\bar{S}_s$ can be explicitly 
given in the following form, where each integral is now unique and we discard
 integrals that do not give any relevant contribution:
\begin{eqnarray}
&&\bar{S}_s= \sqrt{\frac{\alpha'}{2}}\kappa_D 
\epsilon_\mu\bar{\epsilon}_\nu q_\rho \Bigg\{ 
\frac{\alpha'}{2}\sum_{i=1} \bar{C}_i^{[\rho} k_i^{\nu]}\Big( k_i^\mu I_i^{ii}+
\sum_{j\neq i} k_j^\mu I_j^{ii}\Big)+\sum_{i\neq j} \bar{A}_{\{i}^\rho
 \bar{A}_{j\}}^\nu \bigg(C_i^\mu I_{ii}^{ji}+\sqrt{\frac{\alpha'}{2}}
 k_i^\mu I_i^{ij}\bigg) \nonumber\\
&&+\sum_{i\neq j\neq l} \bar{A}_j^\rho \bar{A}_l^\nu\bigg(  C_i^\mu 
I_{ii}^{jl}+\sqrt{\frac{\alpha'}{2}} k_i^\mu I_i^{jl} \bigg) +
\sqrt{\frac{\alpha'}{2}}q_\sigma\Bigg[\sum_{i\neq j}\sum_{i\neq l} 
\frac{1}{2} \bar{A}_{\{i}^\sigma \bar{A}_{l\}}^\nu
 A^\rho_{\{i}A^\mu_{j\}}I_{ij}^{li} -\sqrt{\frac{\alpha'}{2}}\sum_{i\neq j} 
\bar{A}^\sigma_{\{i}\bar{A}^\nu_{j\}}C_j^\rho k_i^\mu I_{ij}^{ij}\nonumber\\
&&-\sqrt{\frac{\alpha'}{2}}\sum_{i\neq j\neq l} \bar{A}_{\{i}^\sigma
 \bar{A}^\nu_{l\}}C_{\{i}^\rho k_{j\}}^\mu I_{ij}^{il}-\sum_{i\neq j} 
 \bar{A}_{\{i}^\sigma \bar{A}^\nu_{j\}}C_j^\rho C_i^\mu I_{jii}^{ij}
-\sum_{i\neq j\neq l} C_j^\rho C_i^\mu\bigg( \bar{A}_{\{i}^\sigma 
\bar{A}^\nu_{l\}} I_{jii}^{il}+\bar{A}_{\{j}^\sigma 
\bar{A}^\nu_{l\}}I^{jl}_{jii}\bigg)\nonumber\\
&&-\frac{\alpha'}{2}\sum_{i\neq j} \bar{C}_i^{[\sigma}k_i^{\nu]}k_i^\mu
 \bar{C}_j^\rho I_{i}^{iij}-\frac{\alpha'}{2}\sum_{i\neq j} 
\bar{C}_j^{[\sigma}k_j^{\nu]}k_i^\mu\bar{C}_i^\rho I_i^{jji}-
\sqrt{\frac{\alpha'}{2}}\sum_{i\neq j\neq l} k_i^\mu\bigg( 
\bar{A}^\sigma_{\{i}\bar{A}_{l\}}^\nu\bar{C}_j^\rho I_i^{jli}+
\bar{A}_l^\sigma \bar{A}_j^\nu \bar{C}_i^\rho I_i^{lji}\bigg)\Bigg]\Bigg\}
\nonumber\\
&&+{\cal O}(q^2)\label{Ss2}
\end{eqnarray}
 where we made explicit use of Eq. (\ref{nulrelation}) and particularly of the
 identity $\bA_i^\rho \bA_i^\nu = 
\sqrt{\frac{\alpha'}{2}}C_i^{[\rho ,} k_i^{\nu]}$. 
 We recall for convenience the notations:
 \seq{
 C_i^{[\rho ,} k_i^{\nu]} &\equiv C_i^\rho k_i^\nu - C_i^\nu k_i^\rho \\
 A_{\{i}^\mu A_{j\}}^\nu &\equiv A_i^\mu A_j^\nu + A_j^\mu A_i^\nu =
 A_i^{[\mu} A_j^{\nu]}
 }
 where the latter equality is due to the Grassmannian nature of the $A_i$.
 
 The $\Ord(q^0)$ part of $\bar{S}_s$ is obtained from the term
 involving $I_{i}^{jl}$ only, since $I_{ii}^{jl}$ does not have a 
$\Ord(q^{-1})$ term, and the only nonzero part reads:
\ea{
\bar{S}_s &=
\kappa_D \epsilon_\mu\bar{\epsilon}_\nu \frac{\alpha'}{2} 
q_\rho\sum_{i\neq j} \big(\bar{A}_j^\rho\bar{A}_i^\nu +
\bar{A}_i^\rho\bar{A}_j^\nu\big)k_i^\mu I_i^{ij}
+{\cal O}(q)
\nonumber \\
&
= \kappa_D \epsilon_\mu
\bar{\epsilon}_\nu \sum_{i\neq j}
\frac{ q_\rho\bar{A}_i^{[\rho}\bar{A}_j^{\nu]} k_i^\mu}{(k_i\cdot q)(
\bar{z}_i-\bar{z}_j)}+{\cal O}(q) \, .
\label{Ssorderq0}
}
It is worth noticing that this expression does not involve any overall $\alpha'$-factor.

Finally we express explicitly the terms of $\Ord(q)$, which after some 
simplifications read:
\ea{
&\bar{S}_s \Big |_{\Ord(q)} =\, \kappa_D \epsilon_\mu \bar{\epsilon}_\nu  
 \sum_{i\neq j}  \Bigg \{
 q_\rho\left(\frac{\alpha'}{2}\right)^{\frac{3}{2}}
\frac{qk_j \bC_i^{[\rho,}k_i^{\nu]}}{\zbzb{i}{j}} 
\left (\frac{k_i^\mu}{qk_i} - \frac{k_j^\mu}{qk_j}\right ) 
 \nonumber \\
 &
 +q_\rho\sqrt{\frac{\alpha'}{2}}
\frac{\bA_{\{i ,}^\rho \bA_{j\} }^\nu}{\zbzb{i}{j}} \sum_{l\neq i} 
\Bigg [
\frac{qk_l}{qk_i} \left (\frac{C_i^\mu}{\zz{i}{l}} +\sqrt{\frac{\alpha'}{2}} 
k_i^\mu \LN{i}{l}^2 \right )
 + 
\left (\frac{C_l^\mu}{\zz{i}{l}} - \sqrt{\frac{\alpha'}{2}}k_l^\mu \LN{i}{l}^2 \right )  
\Bigg ]
\nonumber \\
&
+q_\rho q_\sigma \Bigg[
\left (
\frac{1}{2}
A_{\{i,}^\sigma A_{j\}}^\mu -\sqrt{\frac{\alpha'}{2}} C_{\{i,}^\sigma
 k_{j\}}^\mu\right ) 
\sum_{l\neq i}\frac{\bA_{\{i ,}^\rho \bA_{l\} }^\nu}{qk_i (\zz{i}{j})(\zbzb{i}{l})} 
-\frac{\alpha'}{2}
\frac{\bC_i^{[\sigma,}k_i^{\nu]} \bC_j^\rho}{(\zbzb{i}{j})^2}
\left (\frac{k_j^\mu}{qk_j} - \frac{k_i^\mu}{qk_i}\right ) 
\nonumber \\
&
 -\sqrt{\frac{\alpha'}{2}} \sum_{l\neq i,j} \frac{k_i^\mu 
\left ( \bC_j^\sigma \bA_{\{ i,}^\rho \bA_{l\}}^\nu +
\frac{1}{2} \bC_i^\sigma \bA_{\{ j,}^\rho \bA_{l\}}^\nu 
\right )}{qk_i (\zbzb{i}{j})(\zbzb{i}{l})}
- \sum_{l\neq i} \frac{C_{[i,}^\sigma C_{j]}^\mu \bA_{\{i ,}^\rho 
\bA_{l\} }^\nu}{qk_i(\zz{i}{j})^2 (\zbzb{i}{l})}  
\Bigg] \Bigg\} \, .
\label{completesubsub}
}

\section{Soft action on the lower-point amplitude}
\label{SoftAction}
\setcounter{equation}{0}

In Sec. \ref{SoftExpansion}} 
we have seen that the  $n$-point  string amplitudes 
with all massless external legs  can be written as the convolution integral 
of $M_n^b$ with $M_n^s$. The dependence of $M_n^b$ on the momenta and polarizations is the same as for  the amplitude of  $n$ massless particles  in the bosonic string, which is in turn already known to obey a soft theorem through 
subsubleading order when the soft particle is a graviton or 
dilaton~\cite{DiVecchia:2015oba,DiVecchia:2016amo}, 
{
i.e.
\ea{
M_{n+1}^b = 
M_n \ast S_b =
\left (
\hat{S}_{\rm bos}^{(-1)} + \hat{S}_{\rm bos}^{(0)} + \hat{S}_{\rm bos}^{(1)} \right ) M_n^b + \Ord(q^2) \, ,
\label{generalsubsub1}
}
where
 \sea{
 \hat{S}_{\rm bos}^{(-1)}  =& \, \kappa_D \, \epsilon_{\mu \nu}^S \sum_{i=1} 
\frac{k_i^\mu k_i^\nu}{k_i \cdot q} \, ,
\label{hatSbos-1}
 \\
 \hat{S}_{\rm bos}^{(0)}
=&
 - i \kappa_D \epsilon_{\mu \nu}^S \sum_{i=1}^n \frac{q_\rho 
k_i^\nu  J_i^{\mu \rho}}{k_i \cdot q} \, ,
\label{hatSbos0}
 \\
 \hat{S}_{\rm bos}^{(1)}
=&
- \kappa_D \frac{\epsilon_{\mu \nu}^S}{2} 
\!\sum_{i=1}^n
\!\left [
\frac{q_\rho J_i^{\mu \rho} q_\sigma J_i^{\nu \sigma}}{k_i \cdot q}
\right .
 +
 \frac{q^\mu \eta^{\nu \rho} q^\sigma 
+q^\mu \eta^{\nu \sigma} q^\rho 
- \eta^{\mu\nu} q^\sigma q^\rho }{qk_i}
\mathbf{A}_{i\rho \sigma} 
 \nonumber \\
&\left .  - \alpha' \left (q^\sigma k_{i}^{\nu} \eta^{\rho \mu}+q^\rho k_{i}^{\mu }
\eta^{\sigma \nu} - \eta^{\rho\mu}\eta^{\sigma \nu} (k_i \cdot q) - q^\rho 
q^\sigma \frac{k_{i}^{\mu}k_{i}^{\nu}}{k_i \cdot q} \right )
\Pi_{i\rho \sigma}   \right ] \, ,
\label{hatSbos1}
}
where the different quantities and operators were defined in the introduction,
Eq.~\eqref{Ji}-\eqref{Ai}.
}

In this section we will establish a soft theorem for gravitons and 
dilatons in superstring amplitudes. 
By using the above results for $M_n^b$,
we will do this by showing that also $M_n^s$ satisfies 
similar soft identities. In this way we will crucially see how the 
supersymmetric part cancels the $\alpha'$-terms in the soft theorem 
of the bosonic string, Eq.~\eqref{hatSbos1}, leaving a superstring soft theorem 
free of 
any $\alpha'$-correction
through subsubleading order.
{Let us first notice the trivial leading order result,
\ea{
M_{n+1} = M_n\ast (S_b + S_s + \bar{S}_s)& = M_n \ast S_b + \Ord(q^0)
= (M_n^b \ast M_n^s)\ast S_b + \Ord(q^0)
\nonumber \\
&=
 M_{n+1}^b \ast M_n^s+ \Ord(q^0) = 
\hat{S}_{\rm bos}^{(-1)} \, M_n+ \Ord(q^0) \, 
}
thus at leading order we can trivially identify  $\hat{S}^{(-1)} = \hat{S}_{\rm bos}^{(-1)} $. 
}

In order to identify the superstring soft operator at subleading order, it is useful,  in analogy with the bosonic calculation~\cite{DiVecchia:2015oba}, to make 
 the holomorphic and antiholomorphic sectors completely independent.  
This is achieved by replacing, in the antiholomorphic sector, the  
momentum $k$ of the hard particles with a spurious quantity $\bar{k}$.  
By doing this, the integrand of a closed string amplitude completely factorizes,
at the cost of ${M}_n\equiv M_n(k_i,\epsilon_i,\bar{k}_i,\bar{\epsilon}_i)$ 
only becoming a physical amplitude after identifying 
$\bar{k}$ with ${k}$. 
This, however, leads us to introduce holomorphic angular momentum 
operators, 
\ea{
L_i^{\mu \rho} =i \left( k_i^\mu \frac{\partial}{\partial k_{i\rho}}-k_i^\rho 
\frac{\partial}{\partial k_{i\mu}} \right ) \, ,
\quad
S_i^{\mu \rho} = i \left ( \epsilon_i^\mu \frac{\partial}{\partial 
\epsilon_{i\rho}}-\epsilon_i^\rho \frac{\partial}{\partial \epsilon_{i\mu}} 
\right ) \, ,
}
with similar expressions for the antiholomorphic quantities. The action of 
these operators  on the superkinematical variables, defined in Eq. (\ref{AiBiCi}), 
gives:
\seq{
&(L_i+S_i)^{\mu \rho} A_j^\sigma = i \delta_{ij} \left (\eta^{\sigma \rho} A_i^\mu - 
\eta^{\sigma \mu} A_i^\rho \right ) \, , \quad
(\bar{L}_i+\bar{S}_i)_i^{\mu \rho} \bA_j^\sigma = i \delta_{ij} \left (\eta^{\sigma \rho} 
\bA_i^\mu - \eta^{\sigma \mu} \bA_i^\rho \right ) \, ,
\\
&(L_i+S_i)^{\mu \rho} C_j^\sigma = i \delta_{ij} \left (\eta^{\sigma \rho} C_i^\mu - 
\eta^{\sigma \mu} C_i^\rho \right ) \, , \quad
(\bar{L}_i+\bar{S}_i)^{\mu \rho} \bC_j^\sigma = i \delta_{ij} \left (\eta^{\sigma \rho} 
\bC_i^\mu - \eta^{\sigma \mu} \bC_i^\rho \right ) \, .
\label{JC}
}
From these identities it is straightforward to  show 
a pseudo-soft theorem at subleading order for any soft state (graviton, dilaton, 
Kalb-Ramond) in the following form:
{\ea{
M_{n+1}=&&-i\kappa_D \epsilon_\mu \bar{\epsilon}_\nu \sum_{i=1}^n
\Bigg[\frac{q_\rho {k}_i^\nu(L_i+S_i)^{\mu\rho}}{qk_i}
+\frac{q_\rho k_i^\mu(\bar{L}_i+\bar{S}_i)^{\nu\rho}}{qk_i}\Bigg]M_n(k_i,
\epsilon_i;\bar{k}_i,\bar{\epsilon}_i)\Bigg|_{k=\bar{k}}+{\cal O}(q)
}
This is easiest to see by noting that in the bosonic string the same expression
holds for the $\Ord(q^0)$ part, as shown in Ref.~\cite{DiVecchia:2015oba}, and therefore also for $M_n^b$ as defined in this work,
and since the operator above is linear on $M_n= M_n^b \ast M_n^s$, it needs only to be checked that the operation above on $M_n^s$ reproduces $S_s + \bar{S}_s$ at $\Ord(q^0)$,
given explicitly in Eq.~\eqref{Ssorderq0}.
}

By taking the symmetric, respectively antisymmetric combinations of the above 
expression in the polarization of the soft state, it is possible to turn the above 
pseudo-soft theorem into a physical soft theorem. We postpone the full 
antisymmetric analysis to a future work, and here focus on the symmetric part, which 
reads:
\ea{
(M_{n+1})_S =  - i \kappa_D \epsilon_{\mu \nu}^S
\sum_{i=1}^n \frac{q_\rho k_i^\nu}{qk_i} \left (
L_i + \bar{L}_i + S_i + \bar{S}_i\right )^{\mu \rho} 
M_n(k_i,\epsilon_i;\bar{k}_i,\bar{\epsilon}_i)\Big|_{k=\bar{k}}+{\cal O}(q)
}
where the sub/superscript $S$ is for symmetric and where 
$\epsilon_{\mu\nu}^{S}=\frac{1}{2}(\epsilon_\mu\bar{\epsilon}_\nu+
\epsilon_\nu\bar{\epsilon}_\mu)$. Now using the 
equivalence $(L_i + \bar{L}_i)^{\mu \rho}
M_n(k_i;\bar{k}_i)|_{k=\bar{k}} \equiv L_i^{\mu \rho}M_n^s(k_i)$, we 
can readily set $\bar{k}=k$ and thus get:
\ea{
(M_{n+1})_S& =  - i \kappa_D \epsilon_{\mu \nu}^S
\sum_{i=1}^n \frac{q_\rho k_i^\nu}{qk_i} \left (
L_i + S_i + \bar{S}_i\right )^{\mu \rho} 
M_n(k_i,\epsilon_i,\bar{\epsilon}_i)
+{\cal O}(q)
\nonumber \\
&
=  - i \kappa_D \epsilon_{\mu \nu}^S \sum_{i=1}^n \frac{q_\rho k_i^\nu 
 J_i^{\mu \rho}}{qk_i}  M_n (k_i,\epsilon_i,\bar{\epsilon}_i) +{\cal O}(q) 
\nonumber \\
&
\equiv \hat{S}^{(0)} M_n(k_i,\epsilon_i,\bar{\epsilon}_i) +{\cal O}(q) 
}
where we identified the total angular momentum operator 
$
J_i^{\mu \rho} = L_i^{\mu \rho} + S_i^{\mu \rho} + \bar{S}_i^{\mu \rho}
$, and in the last line we defined the subleading operator $\hat{S}^{(0)}$.
 This result is the well-known subleading soft theorem for the graviton. Here we 
have shown, however, that it also applies to the dilaton, by taking its proper 
polarization tensor, and furthermore that in superstring theory there are no string corrections to the soft operator through this order.
It follows that to the subleading order, the soft theorem for the graviton and 
dilaton in superstring theory is exactly the same as in bosonic string theory.
{Since $\hat{S}^{(0)} = \hat{S}_{\rm bos}^{(0)}$, we could equally well have 
shown this from 
the computation:
\ea{
\hat{S}^{(0)} M_n
=
\hat{S}^{(0)} (M_n^b\ast M_n^s)
&=
(\hat{S}^{(0)} M_n^b) \ast M_n^s + M_n^b \ast (\hat{S}^{(0)} M_n^s)
\nonumber \\
&
=
\left [M_{n}\ast S_b +M_{n}\ast( S_s + \bar{S}_s) \right ]_{\Ord(q^0)}
}
and checking that $\hat{S}^{(0)} M_n^s$ reproduces $S_s + 
\bar{S}_s$ at $\Ord(q^0)$.
}

{
At the subsubleading order we proceed by considering the recently 
established soft theorem 
in the bosonic string Eq.~\eqref{hatSbos1}. 
Let us also recall that the $\alpha'$-terms in Eq.~\eqref{hatSbos1} arise as a 
consequence of gauge invariance together with the fact that the 
three-point amplitude 
in the bosonic string has terms with  higher powers in $\alpha'$. In 
superstring these 
latter terms  are missing  in the three-point amplitude of massless closed states. 
We thus do not expect that the subsubleading soft operator for the superstring 
contains the part proportional to $\alpha'$. We therefore would like to check, as an ansatz,
whether the action
\ea{
- \kappa_D \frac{\epsilon_{\mu \nu}^S}{2} \sum_{i=1}^n
\left [
\frac{q_\rho J_i^{\mu \rho} q_\sigma J_i^{\nu \sigma}}{qk_i}
+ \frac{q^\mu \eta^{\nu \rho} q^\sigma 
+q^\mu \eta^{\nu \sigma} q^\rho 
- \eta^{\mu\nu} q^\sigma q^\rho }{qk_i}
\mathbf{A}_{i\rho \sigma} 
\right ] M_n
\equiv \hat{S}^{(1)} M_n \, ,
\label{supersubsub}
}
reproduces the explicit results derived in the previous section.
Let us first notice that the term involving $J_i^{\mu \rho} J_i^{\nu \sigma}$ is a nonlinear operator. Therefore the above action, decomposed on the $M_n^b$ and $M_n^s$ parts, gives:
\ea{
\hat{S}^{(1)} M_n &= \hat{S}^{(1)} (M_n^b \ast M_n^s)
\nonumber \\
&=(\hat{S}^{(1)} M_n^b) \ast M_n^s +  M_n^b \ast (\hat{S}^{(1)} M_n^s) 
- \kappa_D \, \epsilon_{\mu \nu}^S \, q_\rho q_\sigma  \sum_{i=1}^n
\frac{(J_i^{\mu \rho} M_n^b) \ast (J_i^{\nu \sigma} M_n^s)}{qk_i} \, ,
\label{TS1Mn}
}
and we would like to check whether this reproduces the explicit expressions given for \mbox{$M_n \ast (S_b + S_s + \bar{S}_s)$}.
Since $\hat{S}^{(1)} M_n^b$ does not reproduce fully the complete subsubleading soft behavior of $M_n^b\ast S_b$, it is useful to know explicitly the remaining part, which is simply derived from the action of the $\alpha'$-terms in Eq.~\eqref{hatSbos1}, reading:
\ea{
&
(M_n^b\ast S_b)\Big |_{\Ord(q)} - (\hat{S}^{(1)} M_n^b) 
\nonumber \\
&=
\kappa_D\epsilon_{\mu \nu}^S\frac{\alpha'}{2}\sum_{i=1}^n \left (q^\sigma k_{i}^\nu \eta^{\rho \mu}+q^\rho k_{i}^\mu 
\eta^{\sigma \nu} - \eta^{\rho\mu}\eta^{\sigma \nu} (k_i \cdot q) - q^\rho 
q^\sigma \frac{k_{i}^\mu k_{i}^\nu}{qk_i} \right )
\Pi_{i\rho \sigma}  M_n^b \nonumber \\
&
=M_n^b \ast  \left [\kappa_D \epsilon_{\mu \nu}^S \frac{\alpha'}{2}\sum_{i=1}^n\sum_{j\neq i}
\frac{q_\rho q_\sigma}{qk_i} C_i^{[\mu} k_i^{\rho]}\left ( \frac{C_j^{[\sigma} k_i^{\nu]}}{(\zz{i}{j})^2} +\sqrt{\frac{\alpha'}{2}} \frac{k_j^{[\sigma}k_i^{\nu]}}{\zz{i}{j}} \right ) + \text{c.c}
\right ] \, . 
\label{alphabosonic}
}
}%
We will explicitly show that this part of $S_b$ is exactly cancelled by the 
additional supersymmetric contributions coming from $S_s + \bar{S}_s$.
Having the above expression at hand and the result from 
Ref.~\cite{DiVecchia:2015oba,DiVecchia:2016amo}, we will not 
need to compute the first term
in Eq.~\eqref{TS1Mn} involving $\hat{S}^{(1)} M_n^b$.  
We need only to consider the action of the last two operators 
of Eq.~\eqref{TS1Mn}.
The derivation is straightforward but tedious, and we therefore leave 
it in the appendix.
The result is:
\begin{eqnarray}
&M _n^b& \ast (\hat{S}^{(1)} M_n^s) 
- \kappa_D \, \epsilon_{\mu \nu}^S \, q_\rho q_\sigma  \sum_{i=1}^n
\frac{(J_i^{\mu \rho} M_n^b) \ast (J_i^{\nu \sigma} M_n^s)}{qk_i} 
\nonumber \\
&&
=(M_n^b\ast M_n^s)\ast \kappa_D \epsilon_{\mu \nu}^S 
\sqrt {\frac{\alpha'}{2}}
\Bigg\{  
\nonumber\\[2mm]
&&
q_\rho\Bigg[
\sum_{i\neq j\neq l} 
\bar{A}_i^\rho \bar{A}_j^\nu \bigg(C_l^\mu {I^{(q^0)}}_{ll}^{ij} +
\sqrt{\frac{\alpha'}{2}} k_l^\mu {I^{(q^0)}}_l^{ij} \bigg)
+
\sum_{i\neq j} \bar{A}^\rho_{\{i }\bar{A}^\nu_{j\}}\bigg(
 C_j^\mu {I^{(q^0)}}^{ij}_{jj} +
\sqrt{\frac{\alpha'}{2}}k_i^\mu {I^{(q^0)}}^{ij}_i\bigg)
\Bigg]
\nonumber\\
&&
+\sqrt{\frac{\alpha'}{2}} q_\rho  q_\sigma\Bigg[\sum_{i\neq j}\sum_{l\neq i} 
\frac{1}{2} \bar{A}_{\{l}^\sigma\bar{A}^\nu_{i\}} A^\rho_{\{j} 
A^\mu_{i\}}{I^{(q^{-1})}}^{il}_{ij}
 -\sqrt{\frac{\alpha'}{2}} \sum_{i\neq j\neq l} \bar{A}^\rho_{\{i } 
\bar{A}^\nu_{j\}} k_{\{i}^\mu C_{l\}}^\sigma {I^{(q^{-1})}}_{il}^{ij}\nonumber\\
 && -\sqrt{\frac{\alpha'}{2}} \sum_{i\neq j} \bar{A}^\rho_{\{i } 
\bar{A}^\nu_{j\}}k_i^\mu C_j^\sigma {I^{(q^{-1})}}_{ij}^{ij}-\sum_{i\neq j} 
\bar{A}_{\{i}^\rho \bar{A}^\nu_{j\}}C_i^\mu C_j^\sigma {I^{(q^{-1})}}_{iij}^{ij}
\nonumber\\
 &&- \sum_{i\neq j\neq  l} C_i^\mu C_l^\sigma \bar{A}^\rho_{\{i } 
\bar{A}^\nu_{j\}}{I^{(q^{-1})}}^{ij}_{iil}  - \sum_{i\neq j\neq l} C_i^\mu 
C_l^\sigma\bar{A}^\rho_{\{l } \bar{A}^\nu_{j\}}{I^{(q^{-1})}}_{iil}^{lj}
 - \sqrt{\frac{\alpha'}{2}}\sum_{i\neq j\neq l}  \bar{A}^\rho_{\{j }
 \bar{A}^\nu_{i\}}k_i^\mu \bar{C}_l^\sigma{I^{(q^{-1})}}_i^{ilj}
\nonumber\\
&&
  -\sqrt{\frac{\alpha'}{2}}\sum_{i\neq l\neq j} \bar{A}_j^\rho \bar{A}_l^\nu 
k_i^\mu \bar{C}_i^\sigma {I^{(q^{-1})}}^{ijl}_i+\sum_{i\neq j} \frac{
 C_j^{[\rho }k_j^{\nu]}C_i^{[\mu} k_i^{\sigma]}}{qk_i(z_i-z_j)^2}
\Bigg)\Bigg]\Bigg\}
 +\text{c.c.}
 \label{JJMS1}
 \end{eqnarray}
The derivation in the Appendix involves first computing the action of the 
operators and then rewriting everything in terms of the expressions for the
 integrals $I_{i_1 i_2 \ldots}^{j_1 j_2 \ldots}$, up to the relevant order. 
Therefore we have introduced the superscripts ${(q^a)}$, $a=-1,0$, 
on the $I_{i_1 i_2 \ldots}^{j_1 j_2 \ldots}$, denoting the relevant order in $q$ to
 which the integrals $I_{i_1 i_2 \ldots}^{j_1 j_2 \ldots}$ have been identified. In 
this way we can directly compare this expression with the explicit 
expression in Eq.~\eqref{Ss2} for $S_s$ through $\Ord(q)$. The last term, which 
has not been expressed in terms of $I_{i_1 i_2 \ldots}^{j_1 j_2 \ldots}$, 
is the `left-over' term from this identification procedure.
All the other terms can be matched one-by-one with similar terms in Eq.~\eqref{Ss2}.
In Eq.~\eqref{Ss2} only the terms not involving $A_i$'s remain unmatched.
Specifically we have:
\ea{
&\left [\hat{S}^{(1)} M_n- (\hat{S}^{(1)} M_n^b)\ast M_n^s \right ]
- \left [M_n\ast(S_s + \bar{S}_s) \right]_{S} \Big |_{\Ord(q)}
\nonumber \\
&
=
\kappa_D \epsilon_{\mu \nu}^S
\Bigg [
\frac{\alpha'}{2} q_\rho  q_\sigma\sum_{i\neq j} 
\frac{ C_j^{[\rho }k_j^{\nu]}C_i^{[\mu} k_i^{\sigma]}}{qk_i(z_i-z_j)^2}
-
\left(\frac{\alpha'}{2}\right)^{3/2}q_\rho \sum_{i=1} 
{C}_i^{[\rho} k_i^{\nu]}\Big( k_i^\mu I^i_{ii}+\sum_{j\neq i} k_j^\mu I^j_{ii}\Big)
\nonumber \\
&\qquad
+
\left(\frac{\alpha'}{2}\right)^{2} q_\rho q_\sigma\sum_{i\neq j} 
{C}_i^{[\sigma}k_i^{\nu]}k_i^\mu {C}_j^\rho I^{i}_{iij}
+\left(\frac{\alpha'}{2}\right)^{2} q_\rho q_\sigma\sum_{i\neq j} 
{C}_j^{[\sigma}k_j^{\nu]}k_i^\mu {C}_i^\rho I^i_{jji}
\Bigg ] + \text{c.c.}
\nonumber \\
&=
\kappa_D \epsilon_{\mu \nu}^S
\left [
\frac{\alpha'}{2} q_\rho  q_\sigma\sum_{i\neq j} \frac{ 
C_i^{[\mu }k_i^{\rho]}C_j^{[\sigma} k_i^{\nu]}}{qk_i(z_i-z_j)^2}
-
\left(\frac{\alpha'}{2}\right)^{\frac{3}{2}}\sum_{i=1}^n\sum_{j\neq i} 
q_\rho q_\sigma\frac{C_i^{[\rho} k_i^{\nu]}k_i^{[\mu}k_j^{\sigma]}}{qk_i(z_i-z_j)}
\right ] + \text{c.c} \, ,
\label{rest}
}
where the first part of the left-hand side was identified with Eq.~\eqref{JJMS1}
using Eq.~\eqref{TS1Mn}.
To arrive to the final equality we made use of:
 \begin{eqnarray}
 I^i_{ii}= \sum_{j\neq i}\frac{qk_j}{qk_i(z_i-z_j)}+{\cal O}(q)~~&;&~~
I^j_{ii}= -\frac{1}{z_i-z_j} +{\cal O}(q) \\
I^i_{iij}= -\frac{2}{\alpha'qk_i(z_i-z_j)^2}+{\cal O}(q^0)~~&;&~~I^j_{iij}= 
\frac{2}{\alpha'qk_j(z_i-z_j)^2}+{\cal O}(q^0)
\end{eqnarray}
The right-hand side of Eq.~\eqref{rest} is exactly equal to the  
$\alpha'$-correction in the bosonic string, given in Eq.~\eqref{alphabosonic}. 
Since the soft-behavior of the bosonic part $M_n^b$ is given by 
Eq.~\eqref{generalsubsub1}, which is exactly $\hat{S}^{(1)}$ plus the 
above $\alpha'$ corrections, we arrive at the conclusion that:
\ea{
\hat{S}^{(1)} M_n = M_n \ast( S_b + S_s + \bar{S}_s)_{S}\big |_{\Ord(q)} = 
(M_{n+1})_{S}\big |_{\Ord(q)}
}
This is a subsubleading soft theorem for the graviton and dilaton in the 
supersymmetric string,
with $\hat{S}^{(1)}$ defined in Eq.~\eqref{supersubsub},
and it is simply equal to the field theory result derived in 
Refs.~\cite{Bern:2014vva,DiVecchia:2015jaq}, without further string corrections. 
To be specific, what we have just observed is that the $\alpha'$ corrections 
appearing in $S_b$ are exactly cancelled by the additional supersymmetry parts 
$S_s + \bar{S}_s$.

\section{String corrections in heterotic string from gauge invariance}
\label{heterotic}
\setcounter{equation}{0}

Both in Ref.~\cite{DiVecchia:2016amo} for the bosonic string and in 
this paper for 
the superstring
we have computed the soft behavior through  subsubleading order by explicitly 
performing the string integrals. On the other hand, in 
Ref.~\cite{DiVecchia:2016amo}
 we have also determined the soft behavior, including the string corrections, 
 by imposing  gauge invariance and the fact that the three-point amplitude  
involving
 massless particles already has string corrections. In this section, we extend this 
 second procedure to the heterotic string fixing also in this case the string 
corrections
 at subsubleading order for a soft graviton or dilaton.  It turns out that, as in 
the bosonic
 string, the soft graviton behavior includes string corrections that are, however, 
absent
 for a soft dilaton.
{ This implies that the soft behavior of the dilaton is 
uniquely encoded in an operator universally applicable to all string theories and field theory.}
 
 The basic ingredient is the
three-point amplitude involving gravitons, dilatons and Kalb-Ramond fields
that in the heterotic string is equal to
 \begin{eqnarray}
2\kappa_D \big[\eta^{\mu\mu_i} q^\alpha_i -
\eta^{\mu\alpha_i}q^{\mu_i} +\eta^{\mu_i\alpha}k_i^\mu -
\frac{\alpha'}{2} k_i^\mu q^{\mu_i}q^\alpha_i \big]\big[\eta^{\nu\nu_i} 
q^\beta_i -\eta^{\nu\beta_i} q^{\nu_i} +\eta^{\nu_i\beta_i}k_i^\nu\big] \, ,
\label{3pointampl}
 \end{eqnarray}
where the three particles have the following momenta and polarizations:
$(q, \mu, \nu ),  (k_i ,\mu_i,  \nu_i)$ and $((-q-k_i), \alpha_i, \beta_i)$.
In writing the previous equation we have used momentum conservation
and we have eliminated terms that are zero when we saturate it with the
three polarization vectors.   

The leading term of the scattering amplitude of $(n+1)$ massless particles, 
when one of them becomes soft, is given by the diagram where the soft particle
is attached to the other hard external particles. Since we are only 
interested in  the 
term corresponding to string corrections and of order $q$ in the 
momentum of the soft particle, this pole term is given by,
 \begin{eqnarray}
M_{n+1}^{\mu\nu}(k_1\dots k_n,\,q)\sim -\alpha' \kappa_D  
\sum_{i=1}^n\epsilon^i_{\mu_i}\bar{\epsilon}^i_{\nu_i} 
k_i^\mu k_i^\nu q^{\mu_i} q^{\alpha_i} \eta^{\nu_i\beta_i}\,\frac
{\eta_{\alpha_i r_i}\eta_{\beta_i s_i}}{2k_iq}\, 
M^{r_i s_i}_n(k_i+q) \, .
\label{poletre}
\end{eqnarray}
In order to get a gauge invariant expression we have to add also a term
that is regular in the soft limit $(q\sim 0)$:
\begin{eqnarray}
 M_{n+1}^{\mu\nu}(k_1\dots k_n,\,q) \mid_{\alpha'}= -\alpha' \kappa_D \sum_{i=1}^n 
\frac{k_i^\mu k_i^\nu}{2k_iq} q^{\rho} q^\sigma T_{i \,\rho\sigma}\,\,
 M_n(k_i+q)+N^{\mu\nu} (q, k_i) \, ,
\label{gauginvbt}
 \end{eqnarray}  
 where we also used
 \begin{eqnarray}
T_{i \,\rho \sigma}= \epsilon_{i\,\rho}\frac{\partial}{\partial \epsilon_i^\sigma}
~~;~~M_n (k_i+q) \equiv \epsilon_i^{r_i}  {\bar{\epsilon}}_i^{s_i} M^{r_i s_i}
(k_i +q) \, ,
\label{Trhosigma}
\end{eqnarray} 
and we have omitted to strip off the polarization 
vectors for the other, $j\neq i$, $n-1$
states. Since the pole term is symmetric under the exchange of the 
indices $\mu$ and
$\nu$, gauge invariance can only determine the symmetric part 
of $N^{\mu \nu}$. This is consistent with the fact that gauge invariance does
not fix the term of order $q$ in the soft limit of the Kalb-Ramond field.

Gauge invariance implies:
\begin{eqnarray}
q_\mu M_{n+1}^{\mu\nu} \mid_{\alpha'}= -\frac{\alpha' \kappa_D}{2}  \sum_{i=1}^n 
k_i^\nu   q^\rho q^\sigma
T_{i\,\rho \sigma}M_n(k_i +q)+q_\mu N^{\mu\nu}(q, k_i)=0 \, .
\label{gaurvt}
\end{eqnarray} 
Expanding for small $q$ we get $N^{\mu \nu} (q=0; k_i) =0$ and
\begin{eqnarray}
\frac{\partial}{\partial q_\rho}N^{\mu\nu} +
\frac{\partial }{\partial q_\mu} N^{\rho \nu} = \frac{\alpha' 
\kappa_D}{2}  \sum_{i=1}^n k_i^\nu \left( T^{\mu\rho}_i+
T^{\rho\mu}_i\right)  M_n (k_i) \, .
\label{orderq66}
\end{eqnarray}
Inserting it in Eq. (\ref{gauginvbt}) we get
\begin{eqnarray}
 M_{n+1}^{\mu\nu} \mid_{\alpha'} =&&- \frac{\alpha'\kappa_D}{2} 
\sum_{i=1}^n\frac{k_i^\nu k_i^\mu}{k_iq} q^\rho q^\sigma 
 T_{i\, \rho \sigma} M_n(k_i)
\nonumber \\
&&+
\frac{\alpha' \kappa_D}{8}  \sum_{i=1}^n q_\rho \left[
k_i^\nu  \left(T^{\mu\rho}_i+T^{\rho\mu}_i\right) +
k_i^\mu  \left(T^{\nu\rho}_i+T^{\rho\nu}_i\right) 
\right] M_n\nonumber\\
&&+
\frac{1}{4} q_\rho \left[ \frac{\partial}{\partial q_\rho} N^{\mu\nu}
-\frac{\partial}{\partial q_\mu}N^{\rho\nu} +
 \frac{\partial}{\partial q_\rho} N^{\nu\mu}
-\frac{\partial}{\partial q_\nu}N^{\rho\mu}
\right] \, ,
\label{mn}
\end{eqnarray}
where we have symmetrized under the exchange of  $\nu$ and $\mu$ 
because, as
already observed,  the amplitude has such a symmetry. Imposing   gauge 
invariance
on the index $\nu$; i.e. $q_\nu M_{n+1}^{\mu \nu}=0$, we get the following 
condition:
\ea{
q_\nu   q_\rho \left[\frac{\partial}{\partial q_\rho} N^{\mu\nu}
-\frac{\partial}{\partial q_\mu}N^{\rho\nu}\right] 
 = \alpha'\kappa_D q_\nu q_\rho
 \sum_{i=1}^n \left[k_i^\mu T^{\nu\rho}_i- \frac{1}{2} k_{i}^\nu 
\left( T^{\mu\rho}_i+T^{\rho\mu}_i
 \right)\right] M_n (k_i)
\label{nextcond}
}
that implies
\ea{
& \frac{1}{2} \left[ \left(
\frac{\partial}{\partial q_\rho} N^{\mu\nu}-\frac{\partial}{\partial q_\mu} 
N^{\rho \nu }\right) + \left(
\frac{\partial}{\partial q_\nu} N^{\mu\rho}-\frac{\partial}{\partial q_\mu} 
N^{\nu\rho}\right) \right] \nonumber \\
&= \alpha' \kappa_D
\sum_{i=1}^n
\left[  \frac{1}{2}k_i^\mu(T^{\nu\rho}_i+T^{\rho\nu}_i) -\frac{1}{4} k_i^\nu
  (T^{\mu \rho}_i+T^{\rho \mu}_i)
-\frac{1}{4} k_i^\rho  (T^{\mu \nu}_i+T^{\nu\mu}_i)\right]M_n (k_i) \, .
\label{fgrt5}
}
From the previous relation we can extract the part that is symmetric 
under the exchange of $\mu$ and $\nu$ obtaining
\ea{
& \frac{1}{4} \left[ \left(
\frac{\partial}{\partial q_\rho} N^{\mu\nu}-\frac{\partial}{\partial q_\mu} 
N^{\rho \nu }\right) + \left(
\frac{\partial}{\partial q_\rho} N^{\mu\nu}-\frac{\partial}{\partial q_\nu} 
N^{\mu\rho}\right) \right] \nonumber \\
&= \alpha' \kappa_D
\sum_{i=1}^n
\left[  \frac{1}{8}k_i^\mu(T^{\nu\rho}_i+T^{\rho\nu}_i) +  
  \frac{1}{8}k_i^\nu(T^{\mu\rho}_i+T^{\rho\mu}_i)  
  -\frac{1}{4} k_i^\rho  (T^{\mu \nu}_i+T^{\nu\mu}_i)\right]M_n (k_i) \, ,
\label{uhy7}
}
{which fixes the last part of Eq.~\eqref{mn}.}
{An alternative way of deriving the previous expression 
is by noticing that Eq.~(\ref{orderq66}), together with the symmetric and antisymmetric 
parts of Eq. (\ref{fgrt5}), under the exchange of $\mu$ and $\nu$, actually allow 
to determine the derivative of $N^{\mu \nu}$:
\ea{
\frac{\partial}{\partial q_\rho}  N^{\mu \nu} = \frac{\alpha' \kappa_D}{4} 
\sum_{i=1}^n \left[ k_i^{\mu} (T^{\nu \rho} + T^{\nu \rho} ) +
k_i^{\nu} (T^{\mu \rho} + T^{\mu \rho} ) -
k_i^{\rho} (T^{\mu \nu} + T^{\nu \mu} )     \right]M_n (k_i)\, .
\label{deriN}
}
One can then use this to fix the last part of Eq.~\eqref{mn}, 
equivalent to Eq.~(\ref{uhy7}).
}

{Inserting Eq.~(\ref{uhy7}) in Eq.~(\ref{mn}) }
we finally get the completely fixed string corrections in the case
of the heterotic string:
\ea{
M_{n+1}^{\mu \nu} \mid_{\alpha'} =& - \frac{\alpha' \kappa_D}{4} \sum_{i=1}^{n} \left[ 
\frac{k_i^{\mu} k_i^{\nu}}{k_iq} q^\rho q^\sigma - q^\rho k_i^{\nu} 
\eta^{\mu \sigma} -  q^\rho k_i^{\mu} \eta^{\nu \sigma}
+ (k_i q) \eta^{\mu \sigma} \eta^{\nu \rho}   \right] 
 \left( T^{\rho \sigma}_i + T^{\sigma \rho}_i \right) M_n (k_i) \, ,
\label{softalphacorre}
}
By saturating it with the dilaton polarization  
$ \epsilon_{\mu \nu}^{(D)} =
\eta_{\mu \nu} - q_\mu {\bar{q}}_\nu - q_\nu {\bar{q}}_\mu$ we get
\begin{eqnarray}
&&\epsilon_{\mu \nu}^{(D)} M_{n+1}^{\mu \nu} \mid_{\alpha'} =
 - \frac{\alpha' \kappa_D}{2} \sum_{i=1}^{n} \left[ -2 q^\rho k_i^\sigma
+ (k_i q) \eta^{\rho \sigma} \right]\frac{1}{2} \left( T^{\rho \sigma}_i + 
T^{\sigma \rho}_i 
\right) M_n (k_i) =0 \, ,
\label{dilatonrr}
\end{eqnarray}
which vanishes because of transversality, $(k_i \epsilon_i)=0$,  gauge invariance, $k_i^\sigma \frac{\partial}{\partial \epsilon_{i}^{\sigma}}M_n =0$, and
momentum conservation, $\sum_{i=1}^n k_i = - q$.

In conclusion, as in the bosonic string and in superstring, 
also in the heterotic
string the soft theorem of the dilaton has no $\alpha'$ corrections.

\section{Conclusions and remarks}
\label{Conclusions}
\setcounter{equation}{0}
 
In this paper we have computed superstring amplitudes with  an arbitrary 
number 
of massless external states in the kinematic region where one of the 
massless states 
carries low momentum, be it a graviton, dilaton or a Kalb-Ramond field. 
The soft 
behaviour of the amplitude has been determined through the subsubleading 
order.
When the soft external state is a graviton or a dilaton
 we have further been able 
to identify 
soft operators that, when acting on the amplitude involving only the hard states, reproduce our results, thus demonstrating a soft theorem for these states.

The calculation is an extension of the one done in Ref.~\cite{DiVecchia:2016amo} 
for the bosonic string and despite the much more complicate expressions of the 
amplitudes it requires exactly  the same ingredients and techniques developed 
for the bosonic theory.  

In the case of the  graviton, we have found that the soft operators coincide up 
to  subsubleading order with the ones already identified in the literature 
without any string correction. More specifically, we have shown that the string 
corrections appearing in the bosonic string are exactly cancelled by the additional 
supersymmetric contributions to the amplitude. This result confirms the validity of 
the procedures developed in Ref.~\cite{Bern:2014vva,DiVecchia:2016amo} where  
the soft behaviour is determined via gauge invariance   from the  interaction 
vertices  with three massless closed string states. The absence of string 
corrections in the soft theorem is  a consequence of the absence of  such corrections  in the three-point amplitude  of massless states  in superstring theory.

 { In the case of the dilaton we have found  a universal soft behavior;
 i.e. it is the same in superstring, as well as in heterotic and  
 bosonic string. The universality is a consequence of 
the vanishing of the string corrections to the soft theorem  in all models. 
{It thus also coincide with the field theory result.}
The dilaton soft operator 
contains the generators of scale transformations at subleading order, and the 
special conformal transformations at subsubleading order, as shown in 
Refs.~\cite{DiVecchia:2015jaq,DiVecchia:2016amo}.
Curiously this property is similar to the soft theorem,  derived recently also in 
Ref.~\cite{DiVecchia:2015jaq}, of another scalar known as a dilaton; i.e. the 
Nambu-Goldstone boson of spontaneously broken conformal symmetry. 
Both  dilatons  couple to the trace of the energy momentum tensor,  but 
they obey slightly different soft theorems through the subsubleading order. 
Understanding this difference, as well as understanding the physical origin 
of the string dilaton soft behavior, are indeed problems that deserve further 
studies.}  

\section*{Acknowledgements}
We thank Josh Nohle for useful comments on the dilaton soft theorem, and Oliver Schlotterer for a critical reading of the manuscript.

{
\appendix
\section{Explicit action of the subsubleading soft operator}
\label{appA}
In  this appendix we  compute the action  of the subsubleading  
soft operator given in Eq.~\eqref{supersubsub} on the $n$-point amplitude 
with only hard particles. We denote this operator by $\hat{S}^{(1)}$, i.e.:
\ea{
\hat{S}^{(1)} =
- \kappa_D \frac{\epsilon_{\mu \nu}^S}{2} \sum_{i=1}^n
\left [
\frac{q_\rho J_i^{\mu \rho} q_\sigma J_i^{\nu \sigma}}{qk_i}
+ \frac{q^\mu \eta^{\nu \rho} q^\sigma 
+q^\mu \eta^{\nu \sigma} q^\rho 
- \eta^{\mu\nu} q^\sigma q^\rho }{qk_i}
\mathbf{A}_{i\rho \sigma} 
\right ] \, ,
\label{hatS1}
}
with $J_i$ the total angular momentum operator and 
$\mathbf{A}_i$ given in Eq.~\eqref{Ai}.

We observed in Sec.~\ref{TheAmplitude} that the superstring amplitudes 
with generically $n$-massless states can be decomposed at the integrand 
level into two parts; i.e. $M_n = M_n^b \ast M_n^s$, where one part 
is {related to} to the bosonic string, and  
{the other} part  is a pure superstring contribution.
We can therefore write the action of $\hat{S}^{(1)}$ on $M_n$ as follows:
\ea{
\hat{S}^{(1)} M_n &= \hat{S}^{(1)} (M_n^b \ast M_n^s)
\nonumber \\
&=(\hat{S}^{(1)} M_n^b) \ast M_n^s +  M_n^b \ast (\hat{S}^{(1)} M_n^s) 
- \kappa_D \, \epsilon_{\mu \nu}^S \, q_\rho q_\sigma  \sum_{i=1}^n
\frac{(J_i^{\mu \rho} M_n^b) \ast (J_i^{\nu \sigma} M_n^s)}{qk_i}
\label{S1Mn}
}
The first term, where the soft operator acts on the bosonic string 
amplitude $M_n^b$, has already been determined in 
Ref.~\cite{DiVecchia:2016amo}, and given in 
Eq.~\eqref{generalsubsub1}, for $\alpha' = 0$.
Here we analyze the remaining action of $\hat{S}^{(1)}$ on 
the full superstring amplitude.

The action of the angular momentum operator on $M_n^b$ and 
$M_n^s$, given respectively in Eq.~\eqref{Mnb} and \eqref{Mns}, is 
easily computed and reads:
\ea{
J_i^{\mu \rho} M_n^b =  i M_n^b \ast \sum_{j\neq i=1}^n \left[\frac{\alpha'}{2}
k_i^{[\mu} k_j^{\rho]} \LN{i}{j}^2 
+ 
\left ({\sqrt{\frac{\alpha'}{2}}}\frac{C_{\{i,}^\mu k_{j\}}^\rho - 
C_{\{i,}^\rho k_{j\}}^\mu }{\zz{i}{j}} +\frac{C_i^{[\mu,} 
C_j^{\rho]}}{(\zz{i}{j})^2}+ \text{c.c} \right )
\right ]
}
and 
\ea{
J_i^{\mu \rho} M_n^s =  i M_n^s \ast\sum_{j\neq i=1}^n
\left[ \frac{ A_{\{i}^\rho A_{j\}}^\mu}{z_i-z_j}+\text{c.c.}\right]
}
where the antisymmetric and symmetric combinations of the indices 
are denoted with $k_i^{[\mu}k_j^{\nu]}=k_i^\mu k_j^\nu-k_i^\nu 
k_j^\mu$ and
$k_i^{\{\mu}k_j^{\nu\}}=k_i^\mu k_j^\nu+k_i^\nu k_j^{\mu}$.

Let us consider in Eq.~\eqref{S1Mn} the `mixing' part, which by the 
above formulas can be written as
\ea{
 &-\sum_{i=1}^n
\frac{q_\rho  q_\sigma }{qk_i}
\,  (J_i^{\mu \rho} M_n^b) \ast (J_i^{\nu \sigma} {M}_n^s)= M_n 
\ast\sum_{i=1}^n
\frac{q_\rho  q_\sigma }{qk_i}
\sum_{j\neq i} \frac{\bar{A}_{\{i,}^\rho \bar{A}_{j\}}^\nu}{\bar{z}_i-
\bar{z}_{j}} \sum_{l\neq i} \Bigg[\frac{\alpha'}{2}
k_i^{[\mu ,} k_l^{\sigma]} \LN{i}{l}^2 
\nonumber \\
&
+
\sqrt{\frac{\alpha'}{2}}
\frac{
C_{\{i,}^\mu k_{l\}}^\sigma
}{\zz{i}{l}}
-
\sqrt{\frac{\alpha'}{2}}
\frac{C_{\{i,}^\sigma k_{l\}}^\mu
}{\zz{i}{l}} 
+ 
\frac{C_i^{[\mu,} C_l^{\sigma]}}{(\zz{i}{l})^2} 
+\sqrt{\frac{\alpha'}{2}}
\frac{
\bC_{l}^{[\mu} k_{i}^{\sigma]}
}{\zbzb{i}{l}} 
+ 
\Bigg]+ \text{c.c.} \, ,
\label{JMJM}
}
where we made use of the Grassmannian identity 
$\bA_i^\alpha \bC_i^\beta = 0$, cf. Eq.~\eqref{nulrelation}, to cancel 
some terms.

The idea is now to rewrite every term in terms of the integrals 
$I_{i_1 i_2 \ldots}^{j_1 j_2 \ldots}$ to be able to directly compare 
with the expression in Eq.~\eqref{Ss2}. 
All the identities involving the integrals that we  give in this appendix,  
are obtained starting from   Eqs.~(\ref{1.22}), (\ref{1.23}) and the explicit 
expression of  the master integrals.

Let us consider the terms one by one:

\begin{itemize}
\item
The terms containing the logarithm can be equivalently written as:
\ea{
\sum_{i\neq j}\sum_{l\neq i} &\frac{\alpha'}{2} q_\rho 
\left(\frac{ qk_l}{qk_i}k_i^\mu-k_l^\mu\right)
\left(\frac{A^\rho_{\{i}A^\nu_{j\}}}{z_i-z_j}+\text{c.c}\right)\log|z_i-z_l|^2
\nonumber \\
=&-\sum_{i\neq j\neq l} \frac{\alpha'}{2} k_l^\mu q_\rho 
\frac{A^\rho_{\{i}A^\nu_{j\}}}{z_i-z_j}\log|z_i-z_l|^2
\nonumber \\
&
+\sum_{i\neq j} \frac{\alpha'}{2} k_i^\mu q_\rho 
{\frac{A^\rho_{\{i}A^\nu_{j\}}}{z_i-z_j}
\left(\log|z_i-z_j|^2 +\sum_{i\neq l} \frac{qk_l}{qk_i} \log|z_i-z_l|^2\right)}+
\text{c.c}\nonumber\\
=& \sum_{i\neq l\neq j} \frac{\alpha'}{2} k_l^\mu q_\rho 
A_i^\rho A_j^\nu {I^{(q^0)}}^l_{ij}+\sum_{i\neq j} \frac{\alpha'}{2} 
k_i^\mu q_\rho A^\rho_{\{i}A^\nu_{j\}}{I^{(q^0)}}^i_{ij}+\text{c.c}
}
where we have used Eqs.~(\ref{1.23}), (\ref{A}) and (\ref{IIj}) to identify:
\ea{
{I^{(q^0)}}^l_{ij}= \frac{ \log\frac{|z_j-z_l|^2}{|z_i-z_l|^2}}{z_i-z_j}~~;~~
{{I^{(q^0)}}^i_{ij}=\frac{\log|z_i-z_j|^2}{z_i-z_j} +\sum_{i\neq l} 
\frac{qk_l}{qk_i} \frac{\log|z_i-z_l|^2}{z_i-z_j}}
}
Here $I^{(q^0)}$ denotes soft expansion of the integral $I$ through 
${\cal O}(q^0)$.

\item
The term involving $C_{\{i}^\mu k_{l\}}^\sigma$ in Eq. (\ref{JMJM}) can 
be written as:
\ea{
\sum_{i=1}^n&\sqrt{\frac{\alpha'}{2}}
\frac{q_\rho  q_\sigma }{qk_i}
\sum_{l;j\neq i} \frac{\bar{A}_{\{i,}^\rho \bar{A}_{j\}}^\nu}{\bar{z}_{i}-
\bar{z}_{j}}\frac{C_{\{i}^\mu k_{l\}}^\sigma}{z_i-z_l}
\nonumber \\
= &\sum_{i\neq j\neq l} \sqrt{\frac{\alpha'}{2}}q_\rho C_l^\mu \bar{A}_i^\rho
\bar{A}_j^\nu I_{ll}^{ji} + \sum_{i\neq j}\sqrt{\frac{\alpha'}{2}} q_\rho 
C_i^\mu \bar{A}^\rho_{\{j}\bar{A}^\nu_{i\}}I_{ii}^{ij}+{\cal O}(q^2)
}
where we have used  Eqs.~(\ref{1.22}), (\ref{1.23}) and the master 
integrals to get:
\ea{
\!\!\!{I^{(q^0)}}_{ll}^{ji}=\frac{1}{\bar{z}_i-\bar{z}_j}\left(\frac{1}{z_i-z_l}-
\frac{1}{z_j-z_l}\right)~;~~{I^{(q^0)}}_{ii}^{ij}=\frac{1}{\bar{z}_i-\bar{z}_j}
\left(\sum_{l\neq i}\frac{qk_l}{qk_i(z_i-z_l)}+\frac{1}{z_i-z_j}\right)
}

\item
In the same way the term in Eq.~(\ref{JMJM}) involving 
$C_{\{i}^\sigma k_{l\}}^\mu$  becomes:
\ea{
&-\sum_{i=1}^n\sqrt{\frac{\alpha'}{2}}
\frac{q_\rho  q_\sigma }{qk_i}
\sum_{l;j\neq i} \frac{\bar{A}_{\{i,}^\rho \bar{A}_{j\}}^\nu}{\bar{z}_i-\bar{z}_j}
\frac{C_{\{i}^\sigma k_{l\}}^\mu}{z_i-z_l} \\
&=-  \left(\frac{\alpha'}{2}\right)^{\frac{3}{2}} q_\sigma q_\rho
\left (\sum_{i\neq j\neq l}  k_i^\mu C_l^\sigma\left ( 
\bar{A}_{\{i }^\rho\bar{A}^\nu_{j\}}I_{il}^{ij} +\bar{A}_{\{l }^\rho
\bar{A}^\nu_{j\}}I_{il}^{lj}\right)
+ \sum_{i\neq j} k_i^\mu C_j^\sigma \bar{A}_{\{i}^\rho
\bar{A}^\nu _{j\}}I_{ij}^{ij}
\right )
+{\cal O}(q^2) \nonumber
}
where
\ea{
\!\!\!\!\!\!\! I_{il}^{ij}= \frac{2}{\alpha'qk_i(\bar{z}_i-\bar{z}_j)(z_i-z_l)}+
O(q^0)~;~~
I_{il}^{lj}=-\frac{2}{\alpha'qk_l(\bar{z}_l-\bar{z}_j)(z_i-z_l)}+O(q^0)
}

\item
The term in Eq.~(\ref{JMJM}) involving $C_i^{[\mu} C_l^{\sigma]}$ is  
rewritten in the form:
\ea{
&\sum_{i=1}^n
\frac{q_\rho  q_\sigma }{qk_i}
\sum_{l;j\neq i} \frac{\bar{A}_{\{i,}^\rho 
\bar{A}_{j\}}^\nu}{\bar{z}_i-\bar{z}_j}\frac{C_i^{[\mu } C_l^{\sigma]}}{(z_i-z_l)^2}
 \\
&= -\frac{\alpha'}{2}q_\sigma q_\rho \left ( \sum_{i\neq i\neq  l}  
C_i^\mu C_l^\sigma \bar{A}_{\{i}^\rho \bar{A}^\nu_{j\}}I_{iil}^{ij}
+ \sum_{i\neq i\neq l}  C_i^\mu C_l^\sigma \bar{A}_{\{l}^\rho 
\bar{A}^\nu_{j\}}I_{iil}^{lj}+ \sum_{i\neq j} \bar{A}_{\{i}^\rho
\bar{A}^\nu_{j\}}C_i^\mu C_j^\sigma I^{ij}_{iij} \right )
+{\cal O}(q^2)
\nonumber
}
where we have used the identities:
\ea{
\!\!\!\!I_{iil}^{ij}= -\frac{2}{\alpha' qk_i(z_i-z_l)^2(\bar{z}_i-
\bar{z}_j)}+{\cal O}(q^0)~~;~~ I_{iil}^{lj}=\frac{2}{\alpha' 
qk_l(z_i-z_l)^2(\bar{z}_l-\bar{z}_j)}+{\cal O}(q^0)
\nonumber
} 
\ea{
I_{iij}^{ij}= -\frac{2}{\alpha' (\bar{z}_i-\bar{z}_j)(z_i-z_j)^2}
\left(\frac{1}{qk_i}+\frac{1}{qk_j}\right)+{\cal O}(q^0)
}
 
 \item
 Finally, the term in Eq.~(\ref{JMJM}) involving $\bar{C}_l^{[\mu} 
k_i^{\sigma]}$ can be written as:
\ea{
 &\sum_{i=1}^n\sqrt{\frac{\alpha'}{2}}
 \frac{q_\rho  q_\sigma }{qk_i}
 \sum_{l;j\neq i} \frac{\bar{A}_{\{i,}^\rho \bar{A}_{j\}}^\nu}{\bar{z}_i-\bar{z}_j}
 \frac{\bar{C}_l^{[\mu} k_i^{\sigma]}}{\bar{z}_i-\bar{z}_l}
 \label{A.46}
 \\
 &=- \sum_{i\neq j\neq l} \sqrt{\frac{\alpha'}{2}}q_\rho \frac{\bar{C}_i^\mu \bar{A}_j^\rho \bar{A}_l^\nu}{(\bar{z}_i-\bar{z}_l)(\bar{z}_i-\bar{z}_j)}
 - \sum_{i\neq j\neq l} \left( \frac{\alpha'}{2}\right)^{\frac{3}{2}} q_\rho q_\sigma k_i^\mu \bar{C}_l^\sigma \bar{A}_{\{j}^\rho \bar{A}_{i\}}^\nu I^{ilj}_i+{\cal O}(q^2)
 \nonumber
}
where the following identity was used:
\ea{
I_i^{ilj}= \frac{2}{\alpha' qk_i(\bar{z}_i-\bar{z}_l)(\bar{z}_i-
\bar{z}_j)}+{\cal O}(q^0) \, .
  }
\end{itemize}

Next we consider the `pure' supersymmetric part 
of Eq.~\eqref{S1Mn} and analyze the term:
\ea{
&-  \sum_{i=1}^n
\frac{q_\rho  q_\sigma }{2qk_i}
M_n^b \ast (J_i^{\mu \rho}J_i^{\nu \sigma} {M}_n^s )
\nonumber \\[2mm]
&= - \sum_{i=1}^n
\frac{q_\rho  q_\sigma }{2qk_i}
M_n^b \ast J_i^{\mu \rho} \left [ 
i\left (\sum_{j\neq i} \frac{A_{\{i,}^\sigma A_{j\}}^\nu}{\zz{i}{j}} + 
\text{c.c} \right )
{M}_n^s
\right ]
\nonumber \\[5mm]
&=
 \sum_{i=1}^n
\frac{q_\rho  q_\sigma }{2qk_i}
M_n^b \ast \left [ 
\left (\sum_{j\neq i} \frac{
A_i^\mu A_j^\nu \eta^{\sigma \rho} + A_i^\rho A_j^\sigma \eta^{\mu \nu} -
\eta^{\nu\rho} A_i^\mu A_j^\sigma -\eta^{\sigma\mu}A_i^\rho 
A_j^\nu }{\zz{i}{j}} + \text{c.c} \right )
\right . \nonumber \\
& \quad
 + \left . \sum_{j,l\neq i} \left ( \frac{(A_{\{i,}^\rho 
A_{j\}}^\mu)(A_{\{i,}^\sigma A_{l\}}^\nu)}{(\zz{i}{j})(\zz{i}{l})} + 
\frac{(A_{\{i,}^\rho A_{j\}}^\mu)(\bA_{\{i,}^\sigma 
\bA_{l\}}^\nu)}{(\zz{i}{j})(\zbzb{i}{l})} + \text{c.c.}
\right ) 
\right ] {M}_n^s
\label{JJMns}
}
The first term after the second equality involving $\eta^{\sigma \rho}$ 
vanishes since $q^2=0$, while all the other terms under the same 
parenthesis can be rewritten in terms of the following   differential 
operator acting on the ${ M}_n^s$:
\begin{eqnarray}
&& \sum_{i=1}^n
\frac{q_\rho  q_\sigma }{2qk_i}
M_n^b \ast 
\sum_{j\neq i}\bigg( \frac{
 A_i^\rho A_j^\sigma \eta^{\mu \nu} -  \eta^{\nu\rho} 
A_i^\mu A_j^\sigma -\eta^{\sigma\mu}A_i^\rho 
A_j^\nu      }{z_{i}-z_{j}} + \text{c.c} \bigg)
{M}_n^s\nonumber\\
&&=- M_n^b \ast\sum_{i=1}^n\bigg( \frac{q^\sigma q^\rho \eta^{\nu\mu}-
q^\rho q^\mu\eta^{\nu\sigma}-q^\sigma q^\mu \eta^{\rho\mu}}{2k_iq}\bigg) 
\mathbf{A}_{i\, \rho\sigma}
{M}_n^s
\end{eqnarray}   
This is nothing but the second part of $\hat{S}^{(1)}$ with opposite sign, as 
given in Eq.~\eqref{hatS1}. Thus the two cancel.

The term in Eq.~(\ref{JJMns}) involving four unbarred $A_i$'s gives:
\begin{eqnarray}
&&\sum_{i=1}^n\frac{q_\rho q_\sigma}{2qk_i} M_n^b\ast\sum_{j;l\neq i}
\bigg( \frac{A_i^\rho A_j^\mu   A_i^\sigma A_l^\nu +A_i^\rho A_j^\mu   
A_l^\sigma A_i^\nu+
 A_j^\rho A_i^\mu   A_i^\sigma A_l^\nu  +  A_j^\rho A_i^\mu   A_l^\sigma 
A_i^\nu }{(\zz{i}{j})(\zz{i}{l})}+
\text{c.c.}\bigg){M}_n^s\nonumber\\
&&=M_n\ast\sqrt{\frac{\alpha'}{2}}q_\rho q_\sigma \left(\sum_{i\neq l\neq j} 
\frac{A_j^\rho A_l^\nu C_i^{[\mu}k_i^{\sigma]}}{2qk_i(z_i-z_j)(z_i-z_l)} +
\sqrt{ \frac{\alpha'}{2}} \sum_{i\neq  j}
\frac{C_j^{[\rho}k_j^{\nu]}C_i^{[\mu}k_i^{\sigma]}}{2qk_i(z_i-z_j)^2}+ 
[\mu \leftrightarrow \nu]\right)+\text{c.c.}\nonumber\\
&&=M_n\ast\sqrt{\frac{\alpha'}{2}} q_\rho\left(\sum_{i\neq l\neq j}
\frac{ A_j^\rho A_l^\nu C_i^\mu}{2(z_i-z_j)(z_i-z_l)}-\frac{\alpha'}{4}q_\sigma
 A_j^\rho A_l^\nu k_i^\mu C_i^\sigma I_{ijl}^i 
\right.\nonumber\\
&&
\qquad \qquad\qquad\qquad
\left . +\sqrt{ \frac{\alpha'}{2}}q_\sigma \sum_{i\neq  j}
\frac{C_j^{[\rho}k_j^{\nu]}C_i^{[\mu}k_i^{\sigma]}}{2qk_i(z_i-z_j)^2}+ 
[\mu \leftrightarrow \nu]\right )+\text{c.c.}+{\cal O}(q^2)\label{A.49}
\end{eqnarray}
where we have used Eq.~(\ref{nulrelation}) and the identities:
\begin{eqnarray}
I_{ijl}^i= \frac{2}{\alpha' k_iq(z_i-z_j)(z_i-z_l)}+{\cal O}(q^0)~~;~~
q_\rho q_\sigma A_i^\rho A_i^\sigma=0~~;~~\sum_{i\neq j\neq l} 
\frac{A_i^\mu A_i^\nu(qA_j)(qA_l)}{qk_i(z_i-z_l)(z_i-z_j)}=0\nonumber\\
\end{eqnarray}
The last identity comes out due to the different parity of the numerator 
and denominator  in the exchange of the indices $l$ and $j$. 
We observe that the term involving $A_j^\rho A_l^\nu C_i^\mu$ 
in Eq.~(\ref{A.49}) will cancel the similar term coming from Eq.~(\ref{A.46}).

The last term in Eq.~(\ref{JJMns}) can be equivalently written in the form:
\ea{
\!\! q_\rho q_\sigma\sum_{i\neq j}\sum_{i\neq l} \frac{ 
A^\mu_{\{i}A^\rho_{j\}}\bar{A}^\nu_{\{i}
\bar{A}^\sigma_{l\}}}{2qk_i(z_i-z_j)(\bar{z}_i-\bar{z}_l)}=\frac{\alpha'}{2} 
q_\rho q_\sigma\sum_{i\neq j}\sum_{i\neq l} \frac{1}{2} 
A^\mu_{\{i}A^\rho_{j\}}\bar{A}^\nu_{\{i}\bar{A}^\sigma_{l\}}I_{ij}^{il}
+{\cal O}(q^2)
}
where we have used the identities
\[I_{ij}^{il}= \frac{2}{\alpha' k_iq(z_i-z_j)(\bar{z}_i-\bar{z}_l)}+
{\cal O}(q^0)~~;~~{I_{ij}^{ij}= \frac{2}{\alpha'|z_i-z_j|^2}\left[ 
\frac{1}{k_iq}+\frac{1}{k_jq}\right]+{\cal O}(q^0)}\]
 which follow from Eqs.~(\ref{1.23}) and (\ref{A}).

We can now summarize the result of Eq.~\eqref{S1Mn}.
We are only interested in the second and third part in that expression, 
since we know already the result of $\hat{S}^{(1)} M_n^b$ from 
Ref.~\cite{DiVecchia:2015oba,DiVecchia:2016amo}. In other words, 
cf. Eq.~\eqref{S1Mn}, we have found that
\begin{eqnarray}
&\hat{S}^{(1)} &(M_n^b\ast M_n^s)- (\hat{S}^{(1)} M_n^b)\ast M_n^s 
=(M_n^b\ast M_n^s)\ast \epsilon_{\mu \nu}^S 
\sqrt {\frac{\alpha'}{2}}
\Bigg\{  
\nonumber\\[2mm]
&&
q_\rho\Bigg[
\sum_{i\neq j\neq l} 
\bar{A}_i^\rho \bar{A}_j^\nu \bigg(C_l^\mu {I^{(q^0)}}_{ll}^{ij} +
\sqrt{\frac{\alpha'}{2}} k_l^\mu {I^{(q^0)}}_l^{ij} \bigg)
+
\sum_{i\neq j} \bar{A}^\rho_{\{i }\bar{A}^\nu_{j\}}\bigg( 
C_j^\mu {I^{(q^0)}}^{ij}_{jj} +
\sqrt{\frac{\alpha'}{2}}k_i^\mu {I^{(q^0)}}^{ij}_i\bigg)
\Bigg]
\nonumber\\
&&
+\sqrt{\frac{\alpha'}{2}} q_\rho  q_\sigma\Bigg[\sum_{i\neq j}
\sum_{l\neq i} \frac{1}{2} \bar{A}_{\{l}^\sigma\bar{A}^\nu_{i\}} 
A^\rho_{\{j} A^\mu_{i\}}{I^{(q^{-1})}}^{il}_{ij}
 -\sqrt{\frac{\alpha'}{2}} \sum_{i\neq j\neq l} \bar{A}^\rho_{\{i } 
\bar{A}^\nu_{j\}} k_{\{i}^\mu C_{l\}}^\sigma {I^{(q^{-1})}}_{il}^{ij}\nonumber\\
 && -\sqrt{\frac{\alpha'}{2}} \sum_{i\neq j} \bar{A}^\rho_{\{i } 
\bar{A}^\nu_{j\}}k_i^\mu C_j^\sigma {I^{(q^{-1})}}_{ij}^{ij}-
\sum_{i\neq j} \bar{A}_{\{i}^\rho \bar{A}^\nu_{j\}}C_i^\mu 
C_j^\sigma {I^{(q^{-1})}}_{iij}^{ij}\nonumber\\
 &&- \sum_{i\neq j\neq  l} C_i^\mu C_l^\sigma 
\bar{A}^\rho_{\{i } \bar{A}^\nu_{j\}}{I^{(q^{-1})}}^{ij}_{iil}  - 
\sum_{i\neq j\neq l} C_i^\mu C_l^\sigma\bar{A}^\rho_{\{l } 
\bar{A}^\nu_{j\}}{I^{(q^{-1})}}_{iil}^{lj}
 - \sqrt{\frac{\alpha'}{2}}\sum_{i\neq j\neq l}  
\bar{A}^\rho_{\{j } \bar{A}^\nu_{i\}}k_i^\mu \bar{C}_l^\sigma{I^{(q^{-1})}}_i^{ilj}
\nonumber\\
&&
  -\sqrt{\frac{\alpha'}{2}}\sum_{i\neq l\neq j} \bar{A}_j^\rho 
\bar{A}_l^\nu k_i^\mu \bar{C}_i^\sigma {I^{(q^{-1})}}^{ijl}_i+
\sum_{i\neq j} \frac{ C_j^{[\rho }k_j^{\nu]}
C_i^{[\mu} k_i^{\sigma]}}{qk_i(z_i-z_j)^2}\Bigg)\Bigg]\Bigg\}
 +\text{c.c.}
 \label{JJMS}
 \end{eqnarray}
 }

\end{document}